\RequirePackage{etex}
\documentclass[runningheads]{llncs}
\usepackage{color}
\usepackage[final]{graphicx}
\usepackage{algorithmic}
\usepackage{epsfig}
\usepackage{amssymb}
\usepackage{amsmath}
\usepackage{multicol}
\usepackage{colortbl}
\usepackage{booktabs}
\usepackage{multirow}

\usepackage[utf8]{inputenc}

\usepackage{amsfonts}
\usepackage{graphicx}
\usepackage[table,xcdraw]{xcolor}
\graphicspath{ {images/} }
\usepackage{crypto/crypto}
\usepackage{crypto/protocol}
\usepackage[boxruled,vlined,linesnumbered]{algorithm2e}
\usepackage[n ,advantage ,operators ,sets, adversary,landau,probability,notions,logic,ff,mm,primitives,events,complexity,asymptotics,keys]{cryptocode}
\usepackage{caption}
\usepackage{subcaption}
\usepackage{adjustbox}
\usepackage{placeins}

\usepackage{float}
\usetikzlibrary{matrix,shapes,arrows,positioning,chains,calc}

\usepackage{tikz}
\usetikzlibrary{fit}
\usetikzlibrary{decorations.pathreplacing}
\newlength{\nodedistance}

\usepackage{amsthm} 
\theoremstyle{plain}
\usepackage{hyperref}
\usepackage{centernot}

\usepackage{lscape}

\usepackage{enumitem}
\usepackage{bigdelim}
\usepackage{adjustbox}
\usepackage{bm}
\usepackage{orcidlink}

\usepackage[absolute,overlay]{textpos}
\usepackage{relsize}
\usepackage{bigdelim}
\usepackage{mathtools}
\usepackage{makecell}

\usepackage[]{todonotes}

\SetKwData{Left}{left}\SetKwData{This}{this}\SetKwData{Up}{up}
\SetKwFunction{Union}{Union}\SetKwFunction{FindCompress}{FindCompress}
\SetKwInOut{Input}{Input}\SetKwInOut{Output}{Output}

\SetAlFnt{\scriptsize}

\usepackage{parskip}

\begin{document}

% \title{\begin{tabular}{c} Masking Gaussian Elimination at Arbitrary Order \\ \small with Application to Multivariate- and Code-Based PQC\end{tabular}}
\title{\begin{tabular}{c} Masking Gaussian Elimination at Arbitrary Order\end{tabular} \\ \small with Application to Multivariate- and Code-Based PQC} 

% \title{Masking Gaussian Elimination at Arbitrary Order}
% \subtitle{with Application to Multivariate- and Code-Based PQC}

% \vspace{-30pt}
\author{Quinten Norga\inst{1}\orcidlink{0000-0003-0983-5664} \and Suparna Kundu\inst{1}\orcidlink{0000-0003-4354-852X} \and Uttam Kumar Ojha\inst{2} \and Anindya Ganguly\inst{3} \and Angshuman Karmakar\inst{3}\orcidlink{0000-0003-2594-588X} \and Ingrid Verbauwhede\inst{1}\orcidlink{0000-0002-0879-076X}}

% \authorrunning{Author1 et al.}

\institute{COSIC, KU Leuven, Belgium\\\email{\{firstname\}.\{lastname\}@esat.kuleuven.be} \and Indian Statistical Institute Kolkata, India\footnote{Part of this work was completed while the author was at COSIC, KU Leuven.}\\\email{uttamkumarojha1729@gmail.com} \and Indian Institute of Technology Kanpur, India\\\email{\{anindyag,\,angshuman\}@cse.iitk.ac.in}}

% \vspace{-30pt}
\maketitle

\begin{abstract} 
Digital signature schemes based on multivariate- and code-based hard problems are promising alternatives for lattice-based signature schemes, due to their small signature size. Gaussian Elimination (GE) is a critical operation in the signing procedure of these schemes. In this paper, we provide a masking scheme for GE with back substitution to defend against first- and higher-order attacks. To the best of our knowledge, this work is the first to analyze and propose masking techniques for multivariate- or code-based DS algorithms.

We propose a masked algorithm for transforming a system of linear equations into row-echelon form. This is realized by introducing techniques for efficiently making leading (pivot) elements one while avoiding costly conversions between Boolean and multiplicative masking at all orders. We also propose a technique for efficient masked back substitution, which eventually enables a secure unmasking of the public output. All novel gadgets are proven secure in the $t$-probing model. Additionally, we evaluate the overhead of our countermeasure for several post-quantum candidates and their different security levels at first-, second-, and third-order, including UOV, MAYO, SNOVA, QR-UOV, and MQ-Sign. Notably, the operational cost of first-, second-, and third-order masked GE is 2.3$\times$ higher, and the randomness cost is 1.2$\times$ higher in MAYO compared to UOV for security levels III and V. In contrast, these costs are similar in UOV and MAYO for one version of level I. We also show detailed performance results for masked GE implementations for all three security versions of UOV on the Arm Cortex-M4 and compare them with unmasked results. Our masked implementation targeting UOV parameters has an overhead of factor 15.1$\times$, 15.2$\times$, and 15.4$\times$ compared to the unprotected implementation for NIST security level I, III, and V.
\end{abstract}

\section{Introduction}\label{sec:intro}

The National Institute of Standards and Technology (NIST) published the first set of Post-Quantum Cryptographic (PQC) standards in August 2024 \cite{FIPS203,FIPS204,FIPS205}. 
Three of the four selected cryptographic schemes are based on hard lattice problems. 
To diversify their portfolio and avoid the dependency on a single hard problem, NIST announced another process~\cite{nist2022call} to standardize additional post-quantum Digital Signature (DS) schemes. 
The submitted schemes are designed from various hard problems, such as code-based, hash-based, and multivariate quadratic (MQ) system-based cryptography. 
Recently, NIST announced that 14 out of 40 initial candidates advanced to the second round~\cite{NISTDS_R2}. 
Among the selected submissions, four signatures are from MQ-based cryptography and use the hash-and-sign paradigm. 
These schemes mainly rely on the computational hardness of solving multivariate quadratic systems, a problem known to be NP-complete~\cite{johnson1979computers}. 
The Unbalanced Oil and Vinegar (UOV) signature scheme is one of the oldest and well studied multivariate construction~\cite{originalUOV}. 
% While suffering from a large public key size, signing and verification are fast and the signature size is small.

Gaussian Elimination (GE) is a key component of the signing procedure in many of the schemes selected for the second round of NIST PQC DS on-ramp~\cite{NISTDS_R2}, used for finding the unique solution of a system of linear equations. 
All MQ-based signature schemes, such as (i) UOV~\cite{uov_spec}, (ii) MAYO~\cite{mayo_spec}, (iii) SNOVA~\cite{snova_spec}, (iv) QR-UOV~\cite{furue2023qr}, (v) MQ-Sign~\cite{mq_sign_spec}, (vi) PROV~\cite{prov_spec}, (vii) VOX~\cite{vox_spec}, (viii) TUOV~\cite{tuov_spec}, (ix) VDOO~\cite{vdoo_spec}, and (x) IPRainbow~\cite{iprainbow_spec} rely on GE during signing.
Recently it has also been used in some code-based (CB) signature schemes such as Wave~\cite{wave_spec}.
As the secret key is used during the signing and the GE procedure, it is a potential target for side-channel attacks.

Side-Channel Analysis (SCA) attacks can have severe impacts on cryptographic implementations, and post-quantum schemes are equally vulnerable~\cite{breaks2022profiling,10.1145/3548606.3560579,falconSCA}. 
SCA attacks extract secret data of the mathematically secure cryptographic algorithm from the cryptographic device by observing the computation and its physical behavior. 
Such attacks can be prevented by ensuring that any computation in the cryptographic algorithm is independent of any secret variables. 

Masking~\cite{ChaJutRaoRoh1999} is a provably secure widely used countermeasure of such attacks \cite{BinomialSampling,improvedDilith,HO_saber}. 
Various SCAs have been demonstrated on the UOV-based signature schemes in the literature~\cite{DBLP:journals/tches/AulbachCKSS23,DBLP:journals/fgcs/YiN18,park2018side}. 
However, there is almost no research on countermeasures for MQ and CB digital signature schemes, including UOV, to prevent potential SCAs.
%It balances between optimizing efficiency and enhancing resistance to SCA.
Even more so, no specialized gadgets for the GE operation, a critical and costly component during signing, have been proposed. 

%\noindent\textbf{GE in PQC:} In our initial investigation, we found that Gaussian elimination (GE) is a fundamental component in many of the schemes submitted to the NIST standardization procedure. Its use in multivariate quadratic (MQ) based signature schemes such as XXXXX during the signing procedure is quite well known. Recently it is also being used in some code-based signature schemes such as Wave~\cite{wave_spec}. Since the secret signing key is involved during the GE of these schemes, it is a potential target for side-channel attacks. Masking [x] is a provably secure method for protection against side-channel attacks. Here, the secret element $\{S\}$ is split into multiple components $\{S_i\}$'s such that $\Sigma_i\{S_i\}=\{S\}$ (arithmetic masking) or $\oplus\{S_i\}=\{S\}$ (Boolean masking). Despite the fact that the MQ and code-based cryptography are at least a couple of decades older than lattice-based cryptography, it is surprising that no masking scheme exists for the signing procedure of these schemes. Since GE is the most crucial and time-consuming component of the signing procedure in these schemes, the overhead of the whole masked signing procedure is bottlenecked by the masking of the GE. However, as we are now working with different shares, performing the GE in the masked domain is non-trivial. We have discussed this in Sec.XX.

\noindent
\textbf{Contributions.}
We propose \textit{first- and higher-order} masked algorithms for solving a system of linear equations using (masked) Gaussian elimination with back substitution (\texttt{SecRowEch} \& \texttt{SecBackSub}), a critical component in MQ and CB signature schemes. 
We formally prove their security in the $t$-probing model and analyze the complexity. 
Our techniques and implementations are highly parametrizable and can be extended to other schemes that rely on GE to solve a system of linear equations. We propose masked gadgets for the following sub-operations in Section \ref{sec:maskinguov}:
\begin{itemize}
    \item For efficiently solving the linear system, it needs to be in \textit{row-echelon form}. We propose masked gadgets for making the pivot element non-zero, by securely adding different rows to the pivot-row if it is zero. The \textit{conditional addition} does not reveal the pivot element itself by relying on a secure non-zero check.
    \item Subsequently, the pivot coefficient needs to be reduced to value one. As directly computing its inverse would require unmasking the pivot element, our approach is based on switching \textit{masking representations} for computing its multiplicative inverse. Our approach is efficient as it exploits the multiplicative masking representation for share-wise multiplication.
    \item Finally, we devise an efficient masked gadget for back substituting the linear equations. We propose unmasking the final output \textit{early} and \textit{partitioned} to minimize the amount of masked operations. Intuitively, an output variable is unmasked once it is computed and its public representation used to compute other output variables.
    % Our first-order masked implementation results in a slowdown of a factor 6.2$\times$ compared to the unmasked GE and a factor \textcolor{red}{TODO and TODO} at second- and third-order (Section~\ref{sec:performance}).
    % Its main cost is making the pivot element non-zero and the full reduction to row-echelon form. 
\end{itemize}
We formally prove the $t$-order security of all proposed (sub-)gadgets in the $t$-probing model. 
We apply our techniques on several promising, UOV-based DS schemes (UOV, MAYO, SNOVA, QR-UOV \& MQ-Sign) and compare their operation and randomness cost for masking the GE with back substitution, at first-, second-, and third-order. 
We analyze and show how their parameter choices impact the cost of masking the GE operation in Table~\ref{tab:costcomparison} (see Section~\ref{sec:costcomparison}).
We provide an arbitrary-order masked implementation (Arm M4 C code\footnote{\url{https://github.com/KULeuven-COSIC/Masking-Gaussian-Elimination}}) of our masked GE and evaluate the performance of our methods for UOV parameters on an Arm Cortex-M4 processor. 
Our first-order masked implementations of GE show an overhead of factor 15.1$\times$, $15.2\times$, and 15.4$\times$ compared to the unmasked GE for UOV-I, UOV-III, and UOV-V, respectively. The most expensive steps for masking GE are ensuring the pivot element is non-zero and the full reduction to row-echelon form (see Section~\ref{sec:performance}).

\FloatBarrier

\section{Preliminaries}\label{sec:prelims}
\subsection{Notation}
We denote a finite field with $q$ elements by $\mathbb{F}_q$, with $q$ always a positive integer. 
%Here, $q$ is a power-of-two positive integer. 
%Each element of $\mathbb{F}_q$ is represented by a polynomial over $\mathbb{F}_2$. 
$\mathbb{F}_q^*$ is used to present $\mathbb{F}_q \setminus \{0\}$. 
All the polynomials, vectors, and matrices are defined over $\mathbb{F}_q$ (or $\mathbb{F}_q^*$).
We used lower-case letters to denote field elements/coefficients (e.g., $x$), bold lower-case letters to denote vectors (e.g., $\text{\textbf{b}}$), and bold upper-case letters to denote matrices (e.g., $\text{\pmb{P}}$). 
Please note that all the vectors are in column form and $\text{\pmb{P}}^T$ represents the transposition of a matrix $\text{\pmb{P}}$. 
%$\text{\pmb{0}}_d$ denotes the $d$-dimensional zero vector and $\text{\pmb{I}}_d$ denotes $d$-dimensional identity matrix. 
%$[n]$ denotes index set $\{1, \ldots , n\}$, $||$ denotes string concatenation, 
The assignment of a variable is written as $:=$ and $x \leftarrow A$ denotes sampling an element uniformly random from set $A$ and assignment to variable $x$. 
All logarithms are in base $2$. 
We denote the selection of coordinates in a vector and matrix as $\mathbf{b}{\scriptstyle[j]}$ and $\text{\pmb{P}}{\scriptstyle[j,k]}$. 
The selection of a specific bit of a field element $x$ is denoted by $x^{[i]}$.
A sequence of $n$ variables ($x_1, \cdots, x_n$) (e.g., shares of variable $x$) is represented as ($x_i$)$_{1 \leq i \leq n}$ or in short as ($x_i$) if the sequence length is obvious from context. 
%$\oplus$ denotes the exclusive-or operation (XOR), $\otimes$ denotes a multiplication between two elements of $\mathbb{F}_q$. 
%These operations are extended for the polynomials and vectors by applying them coefficient-wise. 

\subsection{Gaussian Elimination}
% Describe Gaussian elimination
Gaussian elimination, $\pmb{x} \leftarrow \texttt{Gaussian Elimination} (\text{\pmb{A}},\mathbf{b})$, is an old technique to solve a linear system of the form  $\text{\pmb{A}}\mathbf{x}=\mathbf{b}$, where $\text{\pmb{A}} \in \mathbb{F}_q^{m \times m}$ and vector $\mathbf{b} \in \mathbb{F}_q^m$ are given. 
Let us assume that $\text{\pmb{T}}= [\text{\pmb{A}} ~|~ \mathbf{b}]$.
In Line 4 of Algorithm~\ref{alg:ge}, it is attempted to make the pivot element $\text{\pmb{T}}{\scriptstyle [j,j]}$ non-zero by adding the following rows if it is zero. 
%Since all the following rows are added, this requires constant time computations. 
In Line 8, it is checked if the pivot is still a non-zero pivot, and else the computation is aborted as the matrix $\text{\pmb{A}}$ is not invertible in this case (Line 16).
Subsequently, in Line 11, each current row element is multiplied with the inverse of the pivot element, making the pivot $\text{\pmb{T}}{\scriptstyle [j,j]}=1$. 
The loop in Line 13-14 subtracts a scalar multiple of row $j$ from all rows below the current pivot row, making the elements below the pivot zero. 
Finally, to find $\mathbf{x}$, Line 18 back-substitutes the variables into the system of equations.

\begin{algorithm}[tbh]
\DontPrintSemicolon
\KwData{Linear equation $\text{\pmb{A}}\mathbf{x} = \text{\pmb{b}}$, where matrix $\text{\pmb{A}} \in \mathbb{F}_q^{m \times m}$ and vector $\mathbf{b} \in \mathbb{F}_q^m$}
\KwResult{Unique $\mathbf{x} \in \mathbb{F}_q^m$ such that $\text{\pmb{A}}\mathbf{x} = \mathbf{b}$}
\BlankLine
%$r := (q = 256) \text{ ? } 7 : 15$\;\textcolor{blue}{suparna: Do we need r?}\;
$\text{\pmb{T}} := [\text{\pmb{A}} ~|~ \mathbf{b}]$ \tcc*{$\text{\pmb{T}} \in \mathbb{F}_q^{m \times (m + 1)}$}%\tcc*{$\pmb{A}' = [a_{ij1}]$}
\BlankLine
\For{$j = 1$ upto $m$}{
    \texttt{\#\# Try to make pivot $\text{\pmb{T}}{\scriptstyle [j,j]}$ non-zero}\;
    %$stop := (j + r \leq m) ~?~ j + r ~:~ m$\; \textcolor{blue}{suparna: Do we need stop? It is UOV scheme-specific}\;
    \For{$k = j + 1$ upto $m$}{
        \If{$\text{\pmb{T}}{\scriptstyle [j,j]} == 0$}{
        $\text{\pmb{T}}{\scriptstyle [j,j:m+1]} = \text{\pmb{T}}{\scriptstyle [j,j:m+1]} + \text{\pmb{T}}{\scriptstyle [k,j:m+1]}$\;
        }
    }
    \BlankLine
    \texttt{\#\# Check if pivot is non-zero}\;
    \If{$\text{\pmb{T}}{\scriptstyle [j,j]} \neq 0$}{
        \BlankLine
        \texttt{\#\# Multiply row $j$ with the inverse of its pivot}\;
        $p := \text{\pmb{T}}{\scriptstyle [j,j]}^{-1}$\;
        $\text{\pmb{T}}{\scriptstyle [j,j:m+1]} = p\cdot\text{\pmb{T}}{\scriptstyle [j,j:m+1]}$
        \BlankLine
        \texttt{\#\# Subtract scalar multiple of row $j$ from the rows below}\;
        %\tcc{addition of a scalar multiple of one row to another row}\;
        \For{$k = j + 1$ upto $m$}{
            $\text{\pmb{T}}{\scriptstyle [k,j:m+1]} = \text{\pmb{T}}{\scriptstyle [k,j:m+1]} - \text{\pmb{T}}{\scriptstyle [k,j]}\cdot\text{\pmb{T}}{\scriptstyle [j,j:m+1]}$
        }
    }
  \lElse{\Return{$\perp$}}
}
\For{$j = m$ downto $2$}{
  \For{$k = 1$ upto $j$}{
    $\text{\pmb{T}}{\scriptstyle [k,m+1]} = \text{\pmb{T}}{\scriptstyle [k,m+1]} + \text{\pmb{T}}{\scriptstyle [j,k]}\cdot\text{\pmb{T}}{\scriptstyle [j,m+1]}$
  }
}
\Return{$\pmb{x} := \text{\pmb{T}}{\scriptstyle [:,m+1]}$}
\caption{\texttt{Gaussian Elimination}~\cite{uov_spec,DBLP:journals/tches/ChouKY21}}\label{alg:ge}
\end{algorithm}

% \subsubsection{Multivariate-based Digital Signatures}\hfill\\
%Multivariate signatures mainly rely on the computational hardness of solving multivariate quadratic (MQ) systems, a problem known to be NP-complete~\cite{johnson1979computers}. The Unbalanced Oil and Vinegar (UOV) signature scheme is one of the oldest and well studied multivariate construction~\cite{originalUOV}. It uses a multivariate quadratic map that vanishes over a secret subspace. To invert it, the scheme first fixes some variables randomly, thereby converting the quadratic system into a linear one. Gaussian elimination is then applied to solve the resulting linear system.

%In the second round of the NIST signature standardization process~\cite{NISTDS_R2}, three schemes—Mayo~\cite{mayo_spec}, QR-UOV~\cite{furue2023qr}, and SNOVA—adopted the UOV-based technique, in addition to the original UOV signature~\cite{uov_spec}. Each scheme employs Gaussian elimination to invert the quadratic map as part of their signature generation process.

%\todo{need to distinguish here between public and non-public output variants}

%\subsubsection{Code-based Digital Signatures}\hfill\\Code-based assumptions such as syndrome-decoding, 

\subsection{Masking to Thwart SCA Attacks}\label{sec:masking}
% Side-channel attacks extract secret data of the mathematically secure cryptographic algorithm from the cryptographic device by observing the computation and its physical behavior. 
%Some examples are power consumption information and electromagnetic radiation during the execution of the cryptographic algorithm. 
%This side-channel information regarding the secret data is called leakages. 
%Side-channel attacks can be prevented by ensuring that any computation in the cryptographic algorithm is independent of any secret variables. 
%It ensures the resulting leakage is independent of the secret variables, too. 
%This can be realized by integrating masking techniques~\cite{ChaJutRaoRoh1999} with the cryptographic algorithms. 
Masking~\cite{ChaJutRaoRoh1999} is a well-known countermeasure against side-channel attacks. Here the sensitive variables $x$ are split into multiple, randomized shares $(x_1, \ldots, x_n)$. 
As a result, an attacker who does not have access to all shares ($x_i$) does not learn anything about the secret variable $x$.
The relation between $x$ and its shares $(x_1, \ldots, x_n)$ is some group operation that changes depending on the masking methods. 
The most utilized masking method in the literature is Boolean masking, where $x = x_{1} + \cdots + x_{n}$ and the addition is a simple XOR ($\oplus$). 
In this work, we also use multiplicative masking, where $x = x_1 \otimes x_2 \cdots \otimes x_n$ and $\otimes$ denotes a multiplication between two elements of $\mathbb{F}_q$.

To argue about the security level of a masked implementation and its attackers, Ishai et al.~\cite{ishai2003private} introduced the $t$-probing model. 
It assumes an adversary can probe up to $t$ variables during a cryptographic computation. 
An implementation is $t$-probing secure if any $t$ intermediate values leak no information about the unshared secret and thus can be simulated without the knowledge of this secret. 
% Or, protecting against $t$-order attacks requires $n=t+1$ shares. 
In order to simplify the theoretical security analysis of a larger masked algorithm, they can be broken down into smaller functions (i.e., \textit{gadgets}). 
To prove the probing security of the composition of multiple gadgets, several security notions were introduced in~\cite{CCS:BBDFGS16}. 
Below, we recall these security notions as presented in~\cite{BinomialSampling}.

\begin{definition}[$t$-(Strong-)Non-Interference ($t$-(S)NI) security \cite{CCS:BBDFGS16}]
    A gadget with one output sharing and $m_i$ input shares is t-Non-Interference (t-NI) (resp. t-Strong Non-Interference (t-SNI)) secure if any set of at most $t_1$ probes on its internal wires and $t_2$ probes on wires from its output sharings such that $t_1+t_2 \le t$ can be simulated with $t_1 + t_2$ (resp. $t_1$) shares of each of its $m_i$ input sharings.
\end{definition}

%During the conversion of a matrix to row-echelon form, some shared variables are first computed in a masked manner, after which the shares are recombined and the result in made public. 
%More specifically, after a secure computation it is made public wether or not a pivot element is zero, as this requires terminating the entire calculation (no unique solution).
We recall two extensions of the above-mentioned security notions.
Firstly, for the secure unmasking of a shared variable, we rely on the free-$t$-SNI notion, as introduced in \cite{CorSpiShuf}.
It is a stronger notion than $t$-SNI: all output shares, except one, can be perfectly simulated (using fresh randomness). 
As a result, all output shares of a gadget can be simulated using the encoded (e.g. unmasked) value of the output of that gadget.
\begin{definition}[free-$t$-Strong-Non-Interference (free-$t$-SNI) security \cite{CorSpiShuf}]
    \textit{A gadget with output sharing ($b_i$) and $n$ input shares ($m_i$) is free-t-SNI secure if for any set of at most $t_1$ probes on its internal wires such that $t_1 \leq t$, there exists a subset $I$ of input indices with $|I|\leq t_1$, such that the $t_1$ intermediate variables and the output shares $b_{|I}$ can be perfectly simulated from $m_{|I}$, while for any $O \subsetneq [1,n]$ \textbackslash $I$ the output variables in $b_{|O}$ are uniformly and independently distributed, conditioned on the probed variables and $c_{|I}$.}
\end{definition}

Secondly, we rely on the extended $t$-NIo notion from~\cite{barthe2018masking}, which allows for public outputs. 
Here, certain intermediate values in an algorithm are made public and thus are accessible to attackers. As a result, the simulator can also access the distribution of these intermediate values, to ensure successful simulation of the full gadget.  
%A masked implementation should not leak more information than what is released through a public output. 

\begin{definition}[$t$-Non-Interference with public outputs ($t$-NIo) security \cite{barthe2018masking}]
    \textit{A gadget with public output $b$ and $m_i$ input sharings is $t$-Non-Interference with public outputs ($t$-NIo) secure if any set of at most $t_1$ probes on its internal wires such that $t_1 \leq t$ can be simulated with $t_1$ shares of each of its $m_i$ input sharings and $b$.} 
\end{definition}

In Section \ref{sec:maskinguov}, we formally prove our algorithms to be $t$-NI or $t$-SNI secure with $n=t+1$ shares via simulation.

\FloatBarrier

%\input{sections/sensitivity}
%\FloatBarrier

\section{Masked Gaussian Elimination with Back Substitution}\label{sec:maskinguov}
We now describe a method for solving a system of linear equations using GE and back substitution in a masked manner. 
Our approach is generic and can be applied at first- and higher-order. 
The main algorithms are the conversion of a matrix to its row-echelon form (\texttt{SecRowEch}, Algorithm \ref{alg:rowech}) and the back substitution (\texttt{SecBackSub}, Algorithm \ref{alg:backsub}). 

First, in Sections \ref{sec:addmult} - \ref{sec:auxiliary}, we introduce several novel masked gadgets that are used as subroutines in the main algorithms, including:
\begin{itemize}
    \item \texttt{SecCondAdd}: conditional addition of two Boolean shared vectors (rows). This allows us to securely add two rows together, only if the pivot element of the first row is zero, without directly revealing any information about the pivot.
    \item \texttt{SecScalarMult}: multiplication of a (Boolean) shared vector with a multiplicative shared scalar. This allows us to multiply a row with the masked pivot of a different row, without unmasking and revealing the scalar value. 
    %With our approach, we also minimize the amount of mask-type conversions, which are costly.
    \item \texttt{B2Minv}: Boolean to multiplicative inverse mask conversion, which allows us to make a pivot element one by multiplying the row with its inverse, so efficient back substitution can subsequently be performed.
\end{itemize}

All components are put together to achieve fully masked GE with back substitution in Sections \ref{sec:SecRowEch} - \ref{sec:SecBackSub}. 
Table \ref{tab:allgadgets} gives an overview of all the used gadgets in this work, including gadgets from previous works. 
We also include a short description of the computed functionality and the assumed security properties.

\begin{table}
\centering
\caption{Overview of used gadgets in this work, with $n=t+1$ shares.}
\resizebox{\columnwidth}{!}{%
\begin{tabular}{ lllc } 
\toprule
{\bf Algorithm} & {\bf Description} & {\bf Security} & {\bf Reference} \\
\midrule
\texttt{Refresh}  & Refresh of Boolean masking & $t$-NI & \cite{CCS:BBDFGS16,betcorzei18} \& Alg. \ref{alg:refresh}\\
\texttt{StrongRefresh} & Strong refresh of Boolean masking & $t$-SNI & \cite{CCS:BBDFGS16} \& Alg. \ref{alg:strongrefresh}\\
\texttt{FullAdd}    & Refresh and combine Boolean shares & $t$-NI  & \cite{coron2014secure,barthe2018masking} \& Alg. \ref{alg:fullxor} \\
\texttt{B2M}   & Boolean to multiplicative mask conversion & $t$-SNI & \cite{CHES:GenProQui11,mathieu2018mixing} \& Alg. \ref{alg:B2M}\\
\texttt{B2Minv}  & Boolean to multiplicative inverse conversion & $t$-SNI & Algorithm \ref{alg:B2Minv}\\
\texttt{SecMult} & Multiplication of Boolean shares       & $t$-SNI & \cite{ishai2003private,CCS:BBDFGS16} \\
\texttt{SecNonzero} & Nonzero check of Boolean shares      & $t$-SNI & \cite{chen2024masking} \& Alg.\ref{alg:secnonzero} \\
\texttt{SecCondAdd} & Secure conditional addition           & $t$-SNI    & Algorithm \ref{alg:condadd} \\
\texttt{SecScalarMult} & Masked scalar multiplication & $t$-SNI & Algorithm \ref{alg:scalarmult}\\
\texttt{SecMultSub} & Masked multiplication and subtraction & $t$-NI & Algorithm \ref{alg:multadd}\\
\texttt{SecRowEch}  & Matrix conversion to row echelon form & $t$-NIo & Algorithm \ref{alg:rowech}\\
\texttt{SecBackSub}  & Masked back substitution with public output & $t$-NIo & Algorithm \ref{alg:backsub}\\
\bottomrule
\end{tabular}%
}
\label{tab:allgadgets}
\end{table}

\textbf{Methodology. }All novel gadgets are described by a $t$-order algorithm ($n=t+1$ shares) and accompanied with a detailed description.
We also prove the $t$-(S)NI security in the probing model of all algorithms/gadgets.
The proofs are simulation-based: we show how probes on intermediate and output variables in the algorithms can be perfectly simulated with only a limited number of input shares. 
For algorithms that are composed of smaller gadgets, we rely on the $t$-(S)NI properties of the sub-gadgets.
By iterating over all possible intermediate (and output) variables of each sub-gadget, starting at the output and moving to the input of the algorithm, all required probes for simulation are summed.
Crucially, the set of probes required from the input shares of a $t$-SNI gadget is independent from the amount of probes on its output shares.
%As a result, all gadgets are secure against a probing adversary with $t$ probes and can be used in larger compositions.
%\todo{A remaining question/uncertainty is whether an algorithm with loop, if each iteration can be considered independent if they each use a common (shared) input. They are independent, as each loop can be executed in parallel and no operations are happening across iterations, but an input is shared. If not, need (full)refresh each iteration of that input. Thoughts?}

\subsection{Masked Conditional Addition}\label{sec:addmult}
The conditional addition of two Boolean shared row vectors $(\mathbf{x}_i)$ and $(\mathbf{y}_i)$ is described in Algorithm \ref{alg:condadd}.
Depending on the condition, represented by a Boolean shared bit $(b_i)$, the result $\mathbf{s} = \mathbf{x} + \mathbf{y}$ or $\mathbf{s} = \mathbf{x}$ is computed (Line 3) and returned. 
For each coefficient in the vector, the shared term that is added to $(\mathbf{x}{\scriptstyle[j]}_i)$ is computed using the \texttt{SecAND} gadget (Line 2). This gadget can be seen as a specific invocation of the \texttt{SecMult} gadget for $\operatorname{GF}(2)$ (bit-wise logical AND).

\begin{algorithm}[tbh]
\DontPrintSemicolon
\KwData{1. A Boolean sharing $(\mathbf{x}_i)$ of a (row) vector $\mathbf{x} \in \mathbb{F}_q^{l}$. \\ 
\quad \quad \quad 
2. A Boolean sharing $(\mathbf{y}_i)$ of a (row) vector $\mathbf{y} \in \mathbb{F}_q^{l}$. \\
\quad \quad \quad 
3. A Boolean sharing $(b_i)$ of a coefficient (bit) $b$.}
\KwResult{A Boolean sharing $(\mathbf{s}_i)$ of the vector $\mathbf{s}=\mathbf{x} + b\cdot\mathbf{y} \in \mathbb{F}_q^l$}
\BlankLine
%$(\mathbf{b}_i) := \pmb{0}_{m}$\;
\For{$j = 1$ upto $l$}{
  $(\mathbf{a}{\scriptstyle[j]}_i) := \mathtt{SecAND}((\mathbf{y}{\scriptstyle[j]}_i), (b_i^{[w:1]}))$\tcc*{extend $b$ to $w=\lceil\log(q)\rceil$ bits}
  $(\mathbf{s}{\scriptstyle[j]}_i) := (\mathbf{x}{\scriptstyle[j]}_i + \mathbf{a}{\scriptstyle[j]}_i)$\;
  $(\mathbf{s}{\scriptstyle[j]}_i) = \mathtt{StrongRefresh}((\mathbf{s}{\scriptstyle[j]}_i))$
}
\Return{$(\mathbf{s}_i)$} 
\caption{\texttt{SecCondAdd}}\label{alg:condadd}
\end{algorithm}

\subsubsection{Complexity} \hfill\\
Here, we discuss the run-time complexity (number of operations) and randomness complexity of the \texttt{SecCondAdd} operation. We follow the approach proposed in~\cite{CHES:Coron17,BinomialSampling}. We denote the run-time and randomness complexity of an operation \texttt{Operation} by $T_{\mathtt{Operation}}$ and $R_{\mathtt{Operation}}$, respectively. We also assume that the run-time cost of random number generation is unit time and operands are $w=\lceil\log (q)\rceil$ bits wide. 
% The run-time complexity of \texttt{SecCondAdd} is 
\begin{align*}
    T_{\mathtt{SecCondAdd}}(n,l) &= l\cdot(T_{\mathtt{SecAND}}(n)+n+T_{\mathtt{StrongRefresh}}(n))\\
    &=l\cdot(\frac{7n^2-5n}{2}+n+\frac{3n^2-3n}{2})
    =(5n^2-3n)l\,, \\
% \end{align*}
% \begin{align*}
    R_{\mathtt{SecCondAdd}}(n,l,w) &= l\cdot(R_{\mathtt{SecAND}}(n,w)+0+R_{\mathtt{StrongRefresh}}(n,w))\\
    &=l\cdot((\frac{n^2-n}{2}\cdot w)+(\frac{n^2-n}{2}\cdot w)) 
    =(n^2-n) l w\,.
\end{align*}
    
\subsubsection{Security}\hfill \\
We now show Algorithm \ref{alg:condadd} to be $t$-SNI secure with $n=t+1$ shares. 
This means it provides resistance against an adversary with $t$ probes and allows the algorithm to be used in larger compositions.
\begin{lemma}\label{lem:condadd}
The gadget $\mathtt{SecCondAdd}$ (Algorithm \ref{alg:condadd}) is $t$-SNI secure.
\end{lemma}
\noindent\textit{Proof.} 
We first show that a single iteration $j$ is $t$-SNI secure, which is shown in an abstract diagram in Figure \ref{fig:mcondadd}.
Apart from the gadgets listed in Table \ref{tab:allgadgets}, we model the share-wise addition on Line 3 as $t$-NI ($G_2$).
An adversary can probe each gadget ($G_i$) internally or at its output. 
The number of internal and output probes for each gadget is denoted as $t_{G_i}$ and $o_{G_i}$, respectively. 
The total number of probes $t_{A_{\text{\ref{alg:condadd}}}}$ and output shares $|O|$ of (an iteration of) Algorithm \ref{alg:condadd} are:
\begin{equation*}
    t_{A_{\text{\ref{alg:condadd}}}} = \sum\limits_{i=1}^{3}t_{G_i} + \sum\limits_{i=1}^{2}o_{G_i}, \quad |O| = o_{G_3}\,.
\end{equation*}
We show that all internal and output probes can be perfectly simulated with $\leq t_{A_{\text{\ref{alg:condadd}}}}$ input shares. 
Firstly, to simulate the internal and output probes on gadget $G_3$, only $t_{G_3}$ shares of the output of $G_2$ are required.
This is a direct result of the $t$-SNI property of $G_3$: the simulation of a $t$-SNI gadget can be performed independent of the number of probed output shares, stopping the propagation of output probes to the input.
Secondly, the simulation of the probes on gadgets $G_1$-$G_2$ requires $t_{G_1} + t_{G_2} + o_{G_2}$ shares of inputs $(\mathbf{x}_i)$, $(\mathbf{y}_i)$ and $(b_i)$, as $G_2$ is $t$-NI secure.
Due to the $t$-SNI property of gadget $G_1$, $t_{G_1}$ input shares are required to simulate $t_{G_1}$ intermediate probes and $o_{G_1}$ output shares.
Finally, we sum up the required shares of the inputs for the simulation of all gadgets $|I|$. 
As $|I| = t_{G_1} + t_{G_2} + o_{G_2} + t_{G_3} \leq t_{A_{\text{\ref{alg:condadd}}}}$ and independent from $|O|$, iteration $j$ of Algorithm \ref{alg:condadd} is $t$-SNI.
\begin{figure}
    \centering
    % \begin{frame}{}
        % \centering
        \scalebox{0.75}{\setlength{\nodedistance}{4mm}
\begin{tikzpicture}[ % <--- used style are moved here 
node distance = \nodedistance,
data/.style = {},
NIop/.style = {rounded corners, draw=black},
SNIop/.style = {double, double distance=0.5mm, rounded corners, draw=black}]

% INPUTS
\node (I_y) [data]  {($\mathbf{y}{\scriptstyle[j]}_i$)};
\node (I_b) [data, below=\nodedistance of I_y] {($b_i$)};
\node (I_x) [data, below=0.5\nodedistance of I_b]  {($\mathbf{x}{\scriptstyle[j]}_i$)};

\coordinate [below=0.5\nodedistance of I_y] (tmp1);
\coordinate [right=\nodedistance of I_y] (tmp2);
\coordinate [right=1.5\nodedistance of I_b] (tmp3);

% GADGETS
\node (G1_secand) [SNIop, right=3\nodedistance of tmp1] {\begin{tabular}{c} $G_1$ \\ \texttt{SecAND}\end{tabular}};

\node (G2_linear) [NIop, below right=2\nodedistance of G1_secand.north east] {\begin{tabular}{c} $G_2$ \\ $+$\end{tabular}};

\coordinate [right=\nodedistance of G1_secand] (tmp4);
\coordinate [right=6\nodedistance of I_x] (tmp5);

\node (G3_refresh) [SNIop, right=\nodedistance of G2_linear] {\begin{tabular}{c} $G_3$ \\ \texttt{StrongRefresh}\end{tabular}};

% % OUTPUTS
\node (O_s) [data, right=\nodedistance of G3_refresh] {($\mathbf{s}{\scriptstyle[j]}_i$)};

% % CONNECT
\draw[->] (I_y) -- (tmp2) -- (tmp2 |- G1_secand.170) -- (G1_secand.170);
\draw[->] (I_b) -- (tmp3) -- (tmp3 |- G1_secand.190) -- (G1_secand.190);
\draw[->] (G1_secand) -- (tmp4) -- (tmp4 |- G2_linear.160) -- (G2_linear.160);
\draw[->] (I_x) -- (tmp5) -- (tmp5 |- G2_linear.200) -- (G2_linear.200);
\draw[->] (G2_linear) -- (G3_refresh);
\draw[->] (G3_refresh) -- (O_s);

\end{tikzpicture}}
    % \end{frame}
    \caption{An abstract diagram of an iteration $j$ in \texttt{SecCondAdd} (Algorithm \ref{alg:condadd}). 
    The $t$-NI gadgets are depicted with a single border, the $t$-SNI gadgets with a double border.}
    \label{fig:mcondadd}
\end{figure}

Now we remark that each iteration $j$ is independent and can be executed in parallel, meaning that an adversary who places $t$ probes across all iterations can simulate them with no more number of input shares.
All gadgets can thus be summarized into a single gadget (across iterations), meaning the entire loop is $t$-SNI.
\qed

\subsection{Masked Scalar Multiplication}\label{sec:scamult}
The \texttt{SecScalarMult} gadget is described in Algorithm \ref{alg:scalarmult}.
The operation is used to multiply one row-vector ($\mathbf{x}$) of a matrix with a non-zero scalar value ($p$), its pivot-element.
It computes the multiplication $y=p \cdot x$, with one Boolean shared operand $(\mathbf{x}_i)$ and a multiplicative shared operand $(p_i)$.
The end-result is also Boolean shared, as is preferred for succeeding operations.

Our approach does not rely on first converting both operands to the same sharing type. 
The conversion of $(\mathbf{x}_i)$ to the multiplicative domain (\texttt{B2M}) would allow for simple share-wise computation of the multiplication, but would require another conversion back (\texttt{M2B}).
The conversion of $p_i$ to the Boolean domain (\texttt{M2B}) would result in the requirement for \texttt{SecMult} for the multiplication, which has a higher cost than \texttt{SecScalarMult}. %\todo{is this true? should we just leave this as future work}

\begin{algorithm}[tbh]
\DontPrintSemicolon
\KwData{1. A Boolean sharing $(\mathbf{x}_i)$ of vector $x \in \mathbb{F}_q^l$.\\
\quad \quad \quad
2. A multiplicative sharing $(p_i)$ of a coefficient $p \in \mathbb{F}_q$.}
\KwResult{A Boolean sharing $(\mathbf{y}_i)$ of the vector $y=p \cdot x \in \mathbb{F}_q^l$}
\BlankLine
$(\mathbf{y}_i) := (\mathbf{x}_i)$\;
\For{$j=1$ upto $n$}{
    \For{$k=1$ upto $l$}{
        $(\mathbf{y}{\scriptstyle[k]}_i) = (p_j \cdot \mathbf{y}{\scriptstyle[k]}_i)$\;
        $(\mathbf{y}{\scriptstyle[k]}_i) = \mathtt{Refresh}((\mathbf{y}{\scriptstyle[k]}_i))$\;
    }
}
\Return $(\mathbf{y}_i)$
\caption{\texttt{SecScalarMult}}\label{alg:scalarmult}    
\end{algorithm}

We propose to perform the multiplication on one Boolean shared and multiplicative shared operand.
The multiplicative shares are multiplied with all shares of the first operand, one share at a time (Line 2-5).
To ensure no shares of $(p_i)$ are re-combined during the computation, a mask refreshing is used after each multiplication (Line 5).

\subsubsection{Complexity}\hfill\\
% \textcolor{red}{TODO, based on complexity of individual gadgets.}
The run-time and randomness complexity of \texttt{SecScalarMult} are: 
\begin{align*}
    T_{\mathtt{SecScalarMult}}(n,l) &= n\cdot l\cdot (n+T_{\mathtt{Refresh}}(n)) %\\
    %&=n\cdot l\cdot(n + (4n-3))
    =(5n^2-3n)l\,,\\
% \end{align*}
% \begin{align*}
    R_{\mathtt{SecScalarMult}}(n,l,w) &= n\cdot l\cdot (0+R_{\mathtt{Refresh}}(n,w))%\\
    %&=n\cdot l\cdot((n-1)\cdot w) 
    =(n^2-n) l w\,.
\end{align*}
\subsubsection{Security}\hfill \\
We now show that \texttt{SecScalarMult} is $t$-SNI secure with $n=t+1$ shares. 
This means it provides resistance against an adversary with $t$ probes and allows the algorithm to be used in larger compositions.
\begin{lemma}\label{lem:scalarmult}
The gadget $\mathtt{SecScalarMult}$ (Algorithm \ref{alg:scalarmult}) is $t$-SNI secure.
\end{lemma}
\noindent\textit{Proof.} 
Before proving that the entire (outer) loop is $t$-SNI secure (Line 2-5), we show that the inner loop (Line 3-5) is $t$-NI secure, and its operations can be modeled as single $t$-NI gadgets.
This follows directly from each iteration being independent and assumed to be executed in parallel (on individual coefficients). 
As a result, we can summarize the operations in the inner loop into single $t$-NI gadgets $G_1$ (share-wise multiplication) and $G_2$ (\texttt{Refresh}), which operate on the entire vector.

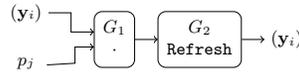
\begin{figure}
    \centering
    % \begin{frame}{}
        % \centering
        \scalebox{0.75}{\setlength{\nodedistance}{4mm}
\begin{tikzpicture}[ % <--- used style are moved here 
node distance = \nodedistance,
data/.style = {},
NIop/.style = {rounded corners, draw=black},
SNIop/.style = {double, double distance=0.5mm, rounded corners, draw=black}]

% INPUTS
\node (I_y) [data]  {($\mathbf{y}_i$)};
\node (I_c) [data, below=\nodedistance of I_y] {$p_j$};

\coordinate [below=0.5\nodedistance of I_y] (tmp1);
\coordinate [right=\nodedistance of I_y] (tmp2);
\coordinate [right=1.5\nodedistance of I_c] (tmp3);

% GADGETS
\node (G1_mult) [NIop, right=3\nodedistance of tmp1] {\begin{tabular}{c} $G_1$ \\ $\cdot$\end{tabular}};

\node (G2_refresh) [NIop, right=\nodedistance of G1_mult] {\begin{tabular}{c} $G_2$ \\ \texttt{Refresh}\end{tabular}};

% % OUTPUTS
\node (O_s) [data, right=\nodedistance of G2_refresh] {($\mathbf{y}_i$)};

% % CONNECT
\draw[->] (I_y) -- (tmp2) -- (tmp2 |- G1_mult.160) -- (G1_mult.160);
\draw[->] (I_b) -- (tmp3) -- (tmp3 |- G1_mult.200) -- (G1_mult.200);
\draw[->] (G1_mult) -- (G2_refresh);
\draw[->] (G2_refresh) -- (O_s);

\end{tikzpicture}}
    % \end{frame}
    \caption{An abstract diagram of an iteration $j$ in \texttt{SecScalarMult} (Algorithm \ref{alg:scalarmult}). 
    The $t$-NI gadgets are depicted with a single border.}
    \label{fig:scalarmult}
% \vspace{-10pt}
\end{figure}

Figure~\ref{fig:scalarmult} depicts an abstract diagram of a single iteration $j$ in Algorithm \ref{alg:scalarmult}.
Since there are $n$ iterations, an adversary cannot place probes in at least one iteration, which we refer to as $j^*$.
We first consider the case where an adversary places all probes after or at the output of iteration $j=j^*$. 
As shown in \cite{betcorzei18}, any set of output shares of \texttt{Refresh} with size $\leq n-1$ is uniformly distributed (Lemma 1).
Thus, all probes can be perfectly simulated with fresh randomness, including the outputs of the entire gadget.

We now show that when an adversary places $t$ probes before $j^*$, these can be simulated with no more number of shares of input $(\mathbf{x}_i)$ and $(p_i)$.
If the share-wise multiplication ($G_1$) and/or refresh ($G_2$) is probed, all probes (across iterations) can be simulated with the same (or fewer) number of input shares of $(\mathbf{x}_i)$.
Now we determine the amount of shares of $(p_i)$ required to simulate all possible probes.
If all $t$ probes are placed in a single iteration $j=j^{'}$, only $p_{j^{'}}$ is required for successful simulation.
If $G_2$ (\texttt{Refresh}) in successive iterations $j=j^{'}-1$ and $j=j^{'}$ are probed, only $p_{j^{'}}$ is required for simulation.
And finally, if probes are placed in non-consecutive iterations of $G_2$, no shares of $(p_i)$ are required. 
The simulation is sound as the outputs of the \texttt{Refresh} gadget are uniformly random, and any such combination of probes can be perfectly simulated with fresh randomness.
We now remark that the size of the required set of shares of $(p_i)$ never exceeds the number of placed probes. 
In conclusion, as the output of the entire loop can be perfectly simulated without any input shares and the required set of input shares of all intermediate probes is less or equal to the amount of placed probes, the entire algorithm is $t$-SNI.
\qed 

\subsection{Masked Multiplication and Subtraction}\label{sec:secmultadd}
The \texttt{SecMultSub} gadget (Algorithm \ref{alg:multadd}) is used to first multiply a Boolean shared (row-)vectors $(\mathbf{x}_i)$ with a Boolean shared coefficient $(c_i)$.
For this, we rely on the \texttt{SecMult} gadget (Line 2).
We do not convert to the multiplicative domain, as the coefficients can be zero and thus would require handling the zero-problem as discussed in \cite{CHES:GenProQui11,mathieu2018mixing}.
The result of the multiplication is subsequently subtracted from a second Boolean shared vector $(\mathbf{y}_i)$ in Line 3.

\begin{algorithm}[tbh]
\DontPrintSemicolon
\KwData{1. Two Boolean sharings $(\mathbf{x}_i), (\mathbf{y}_i)$ of vector $x,y \in \mathbb{F}_q^l$.\\
\quad \quad \quad 
2. A Boolean sharing $(c_i)$ of a coefficient $c \in \mathbb{F}_q$.}
\KwResult{A Boolean sharing $(\mathbf{z}_i)$ of the vector $z= x*c+y \in \mathbb{F}_q^l$}
\BlankLine
\For{$j=1$ upto $l$}{
    $(\mathbf{t}{\scriptstyle[j]}_i) := \mathtt{SecMult}((\mathbf{x}{\scriptstyle[j]}_i), (c_i))$\tcc*{$\mathbf{t}_i \in \mathbb{F}_q^l$}
    $(\mathbf{z}{\scriptstyle[j]}_i) := (\mathbf{y}{\scriptstyle[j]}_i - \mathbf{t}{\scriptstyle[j]}_i)$
}
\Return{$(\mathbf{z}_i)$}
\caption{\texttt{SecMultSub}}\label{alg:multadd}  
\end{algorithm}

\subsubsection{Complexity}\hfill\\
% \textcolor{red}{TODO, based on complexity of individual gadgets.}
The run-time and randomness complexity of \texttt{SecMultAdd} are: 
\begin{align*}
    T_{\mathtt{SecMultAdd}}(n,l) = l\cdot (T_{\mathtt{SecMult}}(n)+n)
    %=l\cdot(\frac{7n^2-5n}{2} + n)
    =\frac{7n^2-3n}{2}\cdot l\,,\\
% \end{align*}
% \begin{align*}
    R_{\mathtt{SecMultAdd}}(n,l,w) = l\cdot (R_{\mathtt{SecMult}}(n,w)+0)
    %=l\cdot(\frac{n^2-n}{2}\cdot w)
    =\frac{n^2-n}{2}\cdot lw\,.
\end{align*}
\subsubsection{Security}\hfill \\
We now show that Algorithm \ref{alg:multadd} is $t$-NI secure with $n=t+1$ shares. 
This means it provides resistance against an adversary with $t$ probes and allows the algorithm to be used in larger compositions.
\begin{lemma}\label{lem:multadd}
The gadget $\mathtt{SecMultSub}$ (Algorithm \ref{alg:multadd}) is $t$-NI secure.
\end{lemma}
\noindent\textit{Proof.} 
Figure \ref{fig:multadd} depicts a single iteration $j$ in \texttt{SecMultSub}.
We model the \texttt{SecMult} gadget as $G_1$ and the share-wise subtraction as $t$-NI gadget $G_2$.
An adversary can probe intermediate values of both gadgets $t_{G_i}$ and the ouput of $G_1$ ($o_{G_1}$).
The total number of adversary probes in an iteration of Algorithm \ref{alg:multadd} is
\begin{equation*}
    t_{A_{\text{\ref{alg:multadd}}}} = \sum\limits_{i=1}^{2}t_{G_i} + o_{G_1}\,.
\end{equation*}
We now show that all probes in iteration $j$ can be simulated with no more number of shares of the inputs ($|I|$) of the iteration: $|I| \leq t_{A_{\text{\ref{alg:multadd}}}}$.
If that is the case, and an adversary can place $t$ probes across different (independent) iterations, those can still be simulated with no more number of input shares.
\begin{figure}[t]
    \centering
    % \begin{frame}{}
        % \centering
        \scalebox{0.75}{\setlength{\nodedistance}{4mm}
\begin{tikzpicture}[ % <--- used style are moved here 
node distance = \nodedistance,
data/.style = {},
NIop/.style = {rounded corners, draw=black},
SNIop/.style = {double, double distance=0.5mm, rounded corners, draw=black}]

% INPUTS
\node (I_y) [data]  {($\mathbf{x}{\scriptstyle[j]}_i$)};
\node (I_c) [data, below=\nodedistance of I_y] {($c_i$)};
\node (I_x) [data, below=0.5\nodedistance of I_b]  {($\mathbf{y}{\scriptstyle[j]}_i$)};

\coordinate [below=0.5\nodedistance of I_y] (tmp1);
\coordinate [right=\nodedistance of I_y] (tmp2);
\coordinate [right=1.5\nodedistance of I_b] (tmp3);

% GADGETS
\node (G1_secand) [SNIop, right=3\nodedistance of tmp1] {\begin{tabular}{c} $G_1$ \\ \texttt{SecMult}\end{tabular}};

\node (G2_linear) [NIop, below right=2\nodedistance of G1_secand.north east] {\begin{tabular}{c} $G_2$ \\ $-$\end{tabular}};

% % OUTPUTS
\node (O_s) [data, right=\nodedistance of G2_linear] {($\mathbf{z}{\scriptstyle[j]}_i$)};

% % CONNECT
\draw[->] (I_y) -- (tmp2) -- (tmp2 |- G1_secand.170) -- (G1_secand.170);
\draw[->] (I_b) -- (tmp3) -- (tmp3 |- G1_secand.190) -- (G1_secand.190);
\draw[->] (G1_secand) -- (tmp4) -- (tmp4 |- G2_linear.160) -- (G2_linear.160);
\draw[->] (I_x) -- (tmp5) -- (tmp5 |- G2_linear.200) -- (G2_linear.200);
\draw[->] (G2_linear) -- (O_s);

\end{tikzpicture}}
    % \end{frame}
    \caption{An abstract diagram of an iteration $j$ in \texttt{SecMultSub} (Algorithm \ref{alg:multadd}). 
    The $t$-NI gadgets are depicted with a single border and the $t$-SNI gadgets with a double border.}
    \label{fig:multadd}
% \vspace{-10pt}
\end{figure}
Due to the $t$-NI property of $G_2$ and $t$-SNI security of $G_1$, it follows directly that $|I| = t_{G_1} + t_{G_2}$, and Lemma \ref{alg:multadd} is proven.
As $t$-SNI security allows to simulate the intermediate and output probes of a gadget with a number of input shares, independent from the amount of probed output shares.
Finally, as each iteration $j$ is independent and can be executed in parallel, we can summarize the gadgets in each iteration as a single gadget across all iterations.
As a result, the entire loop is $t$-NI.
\qed

\subsection{Boolean to Multiplicative Inverse Conversion}\label{sec:B2M}
We now introduce a method for converting from a Boolean to a multiplicative inverse masked representation (Algorithm \ref{alg:B2Minv}).
In order to make a pivot element in a matrix one, as required for it to be in the row-echelon form, we multiply it with its inverse.
As the computation of an inverse prefers multiplicative masking, the Boolean shared input $(x_i)$ is first converted to a multiplicative sharing $(m_i)$ (Line 1).
We rely on the Boolean to multiplicative conversion proposed in \cite{CHES:GenProQui11}, which is recalled in Section \ref{sec:auxiliary}.
Finally, in Line 2, a share-wise inversion is performed.

\begin{algorithm}
\DontPrintSemicolon
\KwData{A Boolean sharing $(x_i)$ of a coefficient $x \in \mathbb{F}_q^*$}
\KwResult{A multiplicative inverse sharing $(p_i)$ such that $x^{-1} = \prod\limits_{i=1}^n p_i$}
\BlankLine
$(m_i) := \mathtt{B2M}((x_i))$\;
$(p_i) := (m_i^{-1})$ \tcc*{multiplicative inverse}
\Return{$(p_i)$}
\caption{\texttt{B2Minv}}\label{alg:B2Minv}  
\end{algorithm}

\subsubsection{Complexity}\hfill \\
% \textcolor{red}{TODO, based on complexity of individual gadgets.}
The run-time and randomness complexity of \texttt{B2Minv} are: 
\begin{align*}
    T_{\mathtt{B2Minv}}(n) = 
    %T_{\mathtt{B2M}}(n)+n=\frac{5n^2-7n+4}{2}+n = 
    \frac{5n^2-5n+4}{2}\,, \text{\quad}
% \end{align*}
% \begin{align*}
    R_{\mathtt{B2Minv}}(n,w) = 
    %R_{\mathtt{B2M}}(n,w)+0=
    \frac{n^2-n}{2}\cdot w\,.
\end{align*}
\subsubsection{Security}\hfill\\
We now show that Algorithm \ref{alg:B2Minv} is $t$-SNI secure with $n=t+1$ shares. 
As a result, it provides resistance against an adversary with $t$ probes and allows the algorithm to be used in larger compositions.
\begin{lemma}\label{lem:b2minv}
The gadget $\mathtt{B2Minv}$ (Algorithm \ref{alg:B2Minv}) is $t$-SNI secure.
\end{lemma}
\noindent\textit{Proof.} 
\texttt{B2M} is a $t$-SNI Boolean to multiplicative conversion, as in \cite{mathieu2018mixing} (Appendix A.2).
The multiplicative inversion on Line 2 can be modeled as a $t$-NI gadget, as it is performed share-wise.
Logically, it follows that Algorithm \ref{alg:B2Minv} is $t$-SNI secure.
\qed

\subsection{Auxiliary Gadgets}\label{sec:auxiliary}
Before discussing our approach to masking the Gaussian elimination and back substitution, we first recall several auxiliary gadgets (see Appendix \ref{sec:appen}). 
We refer to their original work for details on complexity and security.

\texttt{Refresh} \texttt{\&} \texttt{StrongRefresh}. Both types of mask refresh gadgets are used throughout this work and recalled in Algorithm~\ref{alg:refresh} \& \ref{alg:strongrefresh}.
Both were introduced in \cite{CCS:BBDFGS16} (Algorithm 4a \& 4b) and proven to be $t$-NI and $t$-SNI in \cite{betcorzei18} and \cite{CCS:BBDFGS16}, respectively.

\textbf{\texttt{FullAdd}.} We use Algorithm \ref{alg:fullxor} for the secure share recombination.
It consists of two steps: the \textit{strong} mask refreshing and the share combination (unmasking).
In the context of secure unmasking, a strong refresh refers to a free-$t$-SNI mask refreshing.
It is shown in \cite{maskingPolyComp} that the \texttt{StrongRefresh} gadget satisfies the free-$t$-SNI notion.
Thanks to the free-SNI notion, all outputs ($y_i$) are simulatable if the simulator is given the unmasked value $y$. 
As a result, we can recombine the output shares of the gadget (Line 2) while ensuring all intermediate variables can be perfectly simulated.

In contrast, without a free-$t$-SNI gadget in Line 1, the simulation would not be sound. 
Placing an intermediate probe in the unmasking would require all its input shares for simulation. 
A $t$-NI refresh means that to simulate all of its output shares, one would require all input shares, which doesn't allow us to prove the probing security of the full circuit. 
The free-$t$-SNI refresh allows us to simulate all of its output shares (and all intermediate variables of the subsequent unmasking) using all but one of its inputs and the encoded value $y$, which is made public.

As shown in \cite{improvedDilith,maskingPolyComp}, the \texttt{FullAdd} gadget satisfies the $t$-NIo definition when the output $y$ is made public. 
Or, to prove the probing security of a composed circuit, the full gadget can be modeled as $t$-NI if the simulator has knowledge of the encoded output $y$.
%Or, the gadget can be modeled as $t$-NI if it is the final operation in a larger composition, with $y$ a public return value.

\textbf{\texttt{SecNonzero}.} The gadget is recalled in Algorithm \ref{alg:secnonzero}, as introduced in \cite{chen2024masking}.
We refer to the original work for the proof of its $t$-SNI security and Appendix \ref{sec:appen2} for the complexity analysis.
The algorithm checks if a Boolean shared operand $(x_i)$ is non-zero, if unmasked, and returns the result as a single Boolean shared bit $(b_i)$.

\textbf{\texttt{B2M}.} Finally, we also recall the \texttt{B2M} conversion as proposed in \cite{CHES:GenProQui11} (\texttt{AMtoMM}) in Algorithm \ref{alg:B2M}.
Its $t$-SNI security is proven in \cite{mathieu2018mixing} (Appendix A.2).
Intuitively, the algorithm sequentially replaces each Boolean share $x_i$ by a multiplicative one ($z_i$).

\subsection{Masked Row Echelon Conversion}\label{sec:SecRowEch}
A critical step in multivariate-based post-quantum signature schemes, including UOV, is solving a system of linear equations. 
In this (and the next) section, we propose a method for solving a Boolean shared set of linear equations ($(\text{\pmb{T}}_i) = [\text{\pmb{A}}_i ~|~ \mathbf{b}_i]$) using masked Gaussian elimination.
Our strategy consists of reducing a shared matrix, containing the set of equations, to its (masked) row echelon form using (\texttt{SecRowEch}).
Subsequently, solving the system the system requires performing masked back substitution (\texttt{SecBackSub}).

An important step in the computation is the (repeated) checking if the matrix $\text{\pmb{T}}$ is invertible, as this leads to a unique solution. 
We propose an efficient approach, which relies on verifying if pivot elements are (non-)zero in a masked manner.
Note that we do not leak (unmask) the pivot element itself, but securely compute and reveal if it is zero (or not). 
Leaking that the matrix is not invertible is not an issue for security, as the matrix is discarded and the algorithm is re-started in this case.
We now discuss the four steps of gadget \texttt{SecRowEch} in detail.
\begin{algorithm}[tbh]
\DontPrintSemicolon
\KwData{1. A Boolean sharing $(\text{\pmb{A}}_i)$ of matrix $\text{\pmb{A}} \in \mathbb{F}_q^{m \times m}$\\
\quad \quad \quad 
2. A Boolean sharing $(\mathbf{b}_i)$ of the vector $\mathbf{b} \in \mathbb{F}_q^m$} %\\
%\quad \quad \quad
%3. Parameter $q$, $m$}
\KwResult{Masked conversion to row echelon form or $\perp$}
\BlankLine
%$r := (q = 256) \text{ ? } 7 : 15$\;
$(\text{\pmb{T}}_i) := [\text{\pmb{A}}_i ~|~ \mathbf{b}_i]$ \tcc*{$\text{\pmb{T}}_i \in \mathbb{F}_q^{m \times (m + 1)}$}%\tcc*{$\pmb{A}' = [a_{ij1}]$}
\BlankLine
\For{$j = 1$ upto $m$}{
    \texttt{\#\# Try to make pivot ($\text{\pmb{T}}{\scriptstyle [j,j]}$) non-zero}\;
    %$stop := (j + r \leq m) ~?~ j + r ~:~ m$\; 
    \For{$k = j + 1$ upto $m$}{
        $(z_i) := \mathtt{SecNonzero}((\text{\pmb{T}}{\scriptstyle [j,j]}_i))$\;%\tcc*{$b_{ij} \in \{0, 1\}$}
        $(z_i) = \mathtt{SecNOT}((z_i))$\;
        $(\text{\pmb{T}}{\scriptstyle [j,j:m+1]}_i) = \mathtt{SecCondAdd}((\text{\pmb{T}}{\scriptstyle [j,j:m+1]}_i), (\text{\pmb{T}}{\scriptstyle [k,j:m+1]}_i), (z_i))$\;
        % \For{$ l = j$ upto $m + 1$}{
        %     $(\text{\pmb{T}}{\scriptstyle [j,l]}_i) = \mathtt{SecCondAdd}((\text{\pmb{T}}{\scriptstyle [j,l]}_i), (\text{\pmb{T}}{\scriptstyle [k,l]}_i), (z_i))$
        % }
    }
    \BlankLine
    \texttt{\#\# Check if pivot is non-zero}\;
    $(t_i) := \mathtt{SecNonzero}((\text{\pmb{T}}{\scriptstyle [j,j]}_i))$\; %\tcc*{$c_{ij} \in \{0, 1\}$}
    $\mathbf{c}{\scriptstyle [j]} := \mathtt{FullAdd}((t_i))$\;
    \If{$\mathbf{c}{\scriptstyle [j]} \neq 0$}{
        \BlankLine
        \texttt{\#\# Multiply row $j$ with the inverse of its pivot}\;
        $(p_i) := \mathtt{B2Minv}((\text{\pmb{T}}{\scriptstyle [j,j]}_i))$\;
        $(\text{\pmb{T}}{\scriptstyle [j,j:m+1]}_i) = \mathtt{SecScalarMult}((\text{\pmb{T}}{\scriptstyle [j,j:m+1]}_i), (p_i))$
        % \For{$k = j$ upto $m + 1$}{
        %     $(\text{\pmb{T}}{\scriptstyle [j,k]}_i) = \mathtt{SecScalarMult}((\text{\pmb{T}}{\scriptstyle [j,k]}_i), (p_i))$
        % }
        \BlankLine
        \texttt{\#\# Subtract scalar multiple of row $j$ from the rows below}\;
        %\tcc{addition of a scalar multiple of one row to another row}\;
        \For{$k = j + 1$ upto $m$}{
            $(s_i) := \mathtt{StrongRefresh}((\text{\pmb{T}}{\scriptstyle [k,j]}_i))$\;
            $(\text{\pmb{T}}{\scriptstyle [k,j:m+1]}_i) = \mathtt{SecMultSub}((\text{\pmb{T}}{\scriptstyle [j,j:m+1]}_i), (\text{\pmb{T}}{\scriptstyle [k,j:m+1]}_i), (s_i))$
            % \For{$l = j$ upto $m + 1$}{
            %     $(\text{\pmb{T}}{\scriptstyle [k,l]}_i) = \mathtt{SecMultAdd}((\text{\pmb{T}}{\scriptstyle [k,l]}_i), (\text{\pmb{T}}{\scriptstyle [j,l]}_i), (\text{\pmb{T}}{\scriptstyle [k,j]}_i))$
            % }
        }
    }
  \lElse{\Return{$\perp$}}
}
\Return (($\text{\pmb{A}}_i), (\mathbf{b}_i))$
\caption{\texttt{SecRowEch}}\label{alg:rowech}
\end{algorithm}

\textbf{Step 1: try to make pivot $\text{\pmb{T}}{\scriptstyle [j,j]}$ non-zero.} \quad 
As each pivot element (and the rest of the row) is multiplied with its inverse, in order to make it one (row echelon form), it needs to be non-zero.
If the selected pivot element is zero (\texttt{SecNonzero} \& \texttt{SecNOT}\footnote{$(y_i) = $\texttt{SecNOT}$((x_i)) = \neg x_1 + \dots + x_n$}), one of $r$ rows below is added to the `pivot-row' $j$ using \texttt{SecCondAdd} to make it non-zero, all in the masked domain (Line 7).
One needs to iterate over all rows to ensure there are no timing side-channel leakages.

\textbf{Step 2: check if pivot $\text{\pmb{T}}{\scriptstyle [j,j]}$ is non-zero.} \quad 
A masked non-zero check is performed on the pivot element (Line 9), resulting in a Boolean masked bit $(t_i)$.
This value is securely unmasked (\texttt{FullAdd}) and made public in Line 10: if the pivot is still zero the computation is terminated $(c_j==0)$.

\textbf{Step 3: make pivot $\text{\pmb{T}}{\scriptstyle [j,j]} = 1$.} \quad
If the pivot element is non-zero, all elements of its row $(\text{\pmb{T}}{\scriptstyle [j,:]}_i)$ are multiplied with the inverse of the pivot in a masked fashion. 
We rely on the \texttt{B2Minv} gadget: the pivot element is first converted from its Boolean shared form to a multiplicative sharing, which allows us to compute its multiplicative inverse easily (Line 13).
As a result, the \texttt{SecScalarMult} gadget operates on a Boolean shared variable and multiplicative shared scalar (Line 14).
We note that we do not need to represent a zero coefficient in the multiplicative domain (zero-problem, see \cite{CHES:GenProQui11,mathieu2018mixing}) as the computation is aborted before in that case.

\textbf{Step 4: make elements below pivot $\text{\pmb{T}}{\scriptstyle [j,j]}$ zero.} \quad 
Finally, for each of the $k=m-j$ rows below the `pivot-row' (Line 16-18), the element below the pivot is made zero (column $j$). 
Using the \texttt{SecMultSub} gadget in Line 18, the pivot-row $j$ is subtracted $\text{\pmb{T}}{\scriptstyle [k,j]}$ times from each row $k$ below the pivot, in the masked domain.

\subsubsection{Complexity}\hfill \\
The run-time and randomness complexity of \texttt{SecRowEch} are: 
% \begin{align*}
%     T_{\mathtt{SecRowEch}}(n,m) = n\cdot(m^2+m) + \frac{m^2-m}{2}\cdot(T_{\mathtt{SecNonzero}}(n)+T_{\mathtt{SecNOT}}(n)) + \frac{2m^3+3m^2+m}{6}\cdot T_{\mathtt{SecCondAdd}}(n,1) + m\cdot T_{\mathtt{SecNonzero}}(n) + m\cdot T_{\mathtt{FullAdd}}(n) + m + m\cdot T_{\mathtt{B2Minv}}(n) + \frac{m^2+3m}{2}\cdot T_{\mathtt{SecScalarMult}}(n,1) + \frac{m^2-m}{2}\cdot(T_{\mathtt{StrongRefresh}}(n) + \frac{2m^3+3m^2+m}{6}\cdot T_{\mathtt{SecMultAdd}}(n,1)\,,
% \end{align*}
%\textbf{Anindya: Do the calculation}.

\resizebox{.95\linewidth}{!}{
  \begin{minipage}{\linewidth}
\begin{align*}
    T_{\mathtt{SecRowEch}}(n,m) %=& \frac{m^2-m}{2}\cdot(T_{\mathtt{SecNonzero}}(n)+T_{\mathtt{SecNOT}}(n)) + \frac{2m^3+3m^2+m}{6}\cdot \\&T_{\mathtt{SecCondAdd}}(n,1) + m\cdot T_{\mathtt{SecNonzero}}(n) + m\cdot T_{\mathtt{FullAdd}}(n) + m \\&+ m\cdot T_{\mathtt{B2Minv}}(n) + \frac{m^2+3m}{2}\cdot T_{\mathtt{SecScalarMult}}(n,1) + \frac{m^2-m}{2}\cdot\\&T_{\mathtt{StrongRefresh}}(n) + \frac{2m^3+3m^2+m}{6}\cdot T_{\mathtt{SecMultAdd}}(n,1)\\
    =& \frac{m^2-m}{2}\cdot(((5n^2 + 2n - 1)+\lceil\log (w+1)\rceil\cdot(5n^2-n+2))+1) \\&+ \frac{2m^3+3m^2+m}{6}\cdot (5n^2-3n) + m\cdot ((5n^2 + 2n - 1)\\&+\lceil\log (w+1)\rceil\cdot(5n^2-n+2)) + m\cdot \frac{3n^2-n-2}{2} +m \\&+ m\cdot \frac{5n^2-5n+4}{2} + \frac{m^2+3m}{2}\cdot (5n^2-3n) + \frac{m^2-m}{2}\cdot\\&\frac{3n^2-3n}{2} + \frac{2m^3+3m^2+m}{6}\cdot \frac{7n^2-3n}{2}\,,\\
    % =&\,,
    R_{\mathtt{SecRowEch}}(n,m,w) %=& \frac{m^2-m}{2}\cdot(R_{\mathtt{SecNonzero}}(n,w)+R_{\mathtt{SecNOT}}(n,w)) \\&+ \frac{2m^3+3m^2+m}{6}\cdot R_{\mathtt{SecCondAdd}}(n,1,w) + m\cdot \\& R_{\mathtt{SecNonzero}}(n,w) + m\cdot R_{\mathtt{FullAdd}}(n,w) + m\cdot R_{\mathtt{B2Minv}}(n,w) \\& + \frac{m^2+3m}{2}\cdot R_{\mathtt{SecScalarMult}}(n,1,w) + \frac{m^2-m}{2}\cdot \\& R_{\mathtt{StrongRefresh}}(n,w) + \frac{2m^3+3m^2+m}{6}\cdot R_{\mathtt{SecMultAdd}}(n,1,w)\\
    =& \frac{m^2-m}{2}\cdot \frac{\lceil\log (w+1)\rceil^2-\lceil\log (w+1)\rceil}{2}\cdot(n^2-n) \\&+ \frac{2m^3+3m^2+m}{6}\cdot (n^2-n)w + m\cdot \frac{\lceil\log (w+1)\rceil^2 -\lceil\log (w+1)\rceil}{2}\cdot \\&(n^2-n) + m\cdot \frac{(n^2-n)w}{2} + m\cdot \frac{n^2-n}{2}\cdot w + \frac{m^2+3m}{2}\cdot \\& (n^2-n)w + \frac{m^2-m}{2}\cdot (\frac{n^2-n}{2}\cdot w) + \frac{2m^3+3m^2+m}{6}\cdot \\& \frac{n^2-n}{2}\cdot w\,.\\
    % =&(1/12)(6m^3n^2w-6m^3nw+3m^2n^2w^2+15m^2n^2w-3m^2nw^2\\
    % &-15m^2nw+3mn^2w^2+27mn^2w-3mnw^2-27mnw)\\
    % =&\frac{6}{12}m^3n^2w-\frac{6}{12}m^3nw+\frac{3}{12}m^2n^2w^2+\frac{15}{12}m^2n^2w-\frac{3}{12}m^2nw^2\\
    % &-\frac{15}{12}m^2nw+\frac{3}{12}mn^2w^2+\frac{27}{12}mn^2w-\frac{3}{12}mnw^2-\frac{27}{12}mnw\\
    % =& \,.
\end{align*}
\end{minipage}
}

% \resizebox{.95\linewidth}{!}{
%   \begin{minipage}{\linewidth}
% \begin{align*}
%     R_{\mathtt{SecRowEch}}(n,m,w) %=& \frac{m^2-m}{2}\cdot(R_{\mathtt{SecNonzero}}(n,w)+R_{\mathtt{SecNOT}}(n,w)) \\&+ \frac{2m^3+3m^2+m}{6}\cdot R_{\mathtt{SecCondAdd}}(n,1,w) + m\cdot \\& R_{\mathtt{SecNonzero}}(n,w) + m\cdot R_{\mathtt{FullAdd}}(n,w) + m\cdot R_{\mathtt{B2Minv}}(n,w) \\& + \frac{m^2+3m}{2}\cdot R_{\mathtt{SecScalarMult}}(n,1,w) + \frac{m^2-m}{2}\cdot \\& R_{\mathtt{StrongRefresh}}(n,w) + \frac{2m^3+3m^2+m}{6}\cdot R_{\mathtt{SecMultAdd}}(n,1,w)\\
%     =& \frac{m^2-m}{2}\cdot \frac{\lceil\log (w+1)\rceil^2-\lceil\log (w+1)\rceil}{2}\cdot(n^2-n) \\&+ \frac{2m^3+3m^2+m}{6}\cdot (n^2-n)w + m\cdot \frac{\lceil\log (w+1)\rceil^2 -\lceil\log (w+1)\rceil}{2}\cdot \\&(n^2-n) + m\cdot \frac{(n^2-n)w}{2} + m\cdot \frac{n^2-n}{2}\cdot w + \frac{m^2+3m}{2}\cdot \\& (n^2-n)w + \frac{m^2-m}{2}\cdot (\frac{n^2-n}{2}\cdot w) + \frac{2m^3+3m^2+m}{6}\cdot \\& \frac{n^2-n}{2}\cdot w\,.\\
%     % =&(1/12)(6m^3n^2w-6m^3nw+3m^2n^2w^2+15m^2n^2w-3m^2nw^2\\
%     % &-15m^2nw+3mn^2w^2+27mn^2w-3mnw^2-27mnw)\\
%     % =&\frac{6}{12}m^3n^2w-\frac{6}{12}m^3nw+\frac{3}{12}m^2n^2w^2+\frac{15}{12}m^2n^2w-\frac{3}{12}m^2nw^2\\
%     % &-\frac{15}{12}m^2nw+\frac{3}{12}mn^2w^2+\frac{27}{12}mn^2w-\frac{3}{12}mnw^2-\frac{27}{12}mnw\\
%     % =& \,.
% \end{align*}
% \end{minipage}
% }

More details on the computation are provided in Appendix \ref{sec:appen3}.
\subsubsection{Security}\hfill \\
We now argue about the first- and high-order security of Algorithm \ref{alg:rowech} by proving it to be $t$-NIo secure with $n=t+1$ shares and public output $c$. 
This means it provides resistance against an adversary with $t$ probes and allows the algorithm to be used in larger compositions.

\begin{figure}
    \centering
    % \begin{frame}{}
        % \centering
        \scalebox{0.8}{\setlength{\nodedistance}{4mm}
\begin{tikzpicture}[ % <--- used style are moved here 
node distance = \nodedistance,
data/.style = {},
NIop/.style = {rounded corners, draw=black},
SNIop/.style = {double, double distance=0.5mm, rounded corners, draw=black}]

% INPUTS
\node (I_T) [data]  {$(\text{\pmb{T}}_i)$};

\coordinate [right=\nodedistance of I_T] (tmp1);
\coordinate [above=1.3\nodedistance of tmp1] (tmp3);

% GADGETS
\node (step1) [SNIop, right=\nodedistance of tmp1] {\begin{tabular}{c} \texttt{Step 1}\end{tabular}};

\coordinate [right=\nodedistance of step1] (tmp2);
\coordinate [below=2\nodedistance of tmp2] (tmp4);

\node (step2) [NIop, right=\nodedistance of tmp4] {\begin{tabular}{c} \texttt{Step 2}\end{tabular}};

\node (step3) [SNIop, right=\nodedistance of tmp2] {\begin{tabular}{c} \texttt{Step 3}\end{tabular}};

\node (step4) [NIop, above right=0.3\nodedistance and 2\nodedistance of step3.east] {\begin{tabular}{c} \texttt{Step 4}\end{tabular}};

\coordinate [right=\nodedistance of step3] (tmp5);

% OUTPUTS
\node (O_cj) [data, right=\nodedistance of step2]  {$c{\scriptstyle [j]}$};

\node (O_T) [data, right=\nodedistance of step4]  {$(\text{\pmb{T}}_i)$};

% CONNECT
\draw [->] (I_T) -- (tmp1) -- (step1);
\draw [->] (tmp2) -- (tmp4) -- (step2);
\draw [->] (step2) -- (O_cj);
\draw [->] (step1) -- (step3);
\draw [->] (step3) -- (tmp5) -- (tmp5 |- step4.190) -- (step4.190);
\draw [->] (tmp1) -- (tmp3) -- (step4.170);
\draw [->] (step4) -- (O_T);

\end{tikzpicture}}
    % \end{frame}
    \caption{An abstract diagram of a single iteration $j$ of \texttt{SecRowEch} (Algorithm \ref{alg:rowech}). 
    The $t$-NI gadgets are depicted with a single border and the $t$-SNI gadgets with a double border.}
    \label{fig:secrowech}
% \vspace{-10pt}
\end{figure}
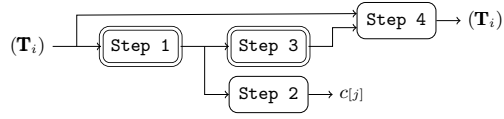

\begin{lemma}\label{lem:SecRowEch}
The gadget $\mathtt{SecRowEch}$ (Algorithm \ref{alg:rowech}) is $t$-NIo secure with public output $\mathbf{c}$.
\end{lemma}
\noindent\textit{Proof.} 
We prove the security of Steps 1 through 4 from their composition of smaller gadgets in Appendix \ref{sec:appen4}.
Step 1 and 3 can be modeled as $t$-SNI gadgets, while Step 2 is $t$-NIo with public output $\mathbf{c}{\scriptstyle [j]}$ and Step 4 is $t$-NI (Figure \ref{fig:secrowech}).
We now prove the larger composition (Algorithm \ref{alg:rowech}) to be $t$-NIo.
An abstract diagram of an iteration $j$ is shown in Figure \ref{fig:secrowech}, constructed from Step 1 - 4.
We now show that a single (outer) loop is $t$-NI secure if the value $\mathbf{c}{\scriptstyle [j]}$ is given to the simulator. 
% An adversary can probe intermediate values $t_{S_i}$ in each step and the output shares of each gadget $o_{S_i}$, except the output of the full iteration.
% The total number of probes in an iteration $j$ is defined as $t_{A_{\text{\ref{alg:rowech}}}}$, with
% \begin{equation*}
%     t_{A_{\text{\ref{alg:rowech}}}} = \sum\limits_{i=1}^{4}t_{S_i} + \sum\limits_{i=0}^{3}o_{S_i}
% \end{equation*}
% We now show that all probes in iteration $j$ can be perfectly simulated with $\leq t_{A_{\text{\ref{alg:rowech}}}}$ shares of the input.
This is a direct result of each step achieving at least $t$-NI security and being composed as in Figure \ref{fig:secrowech} and all public outputs being securely recombined. 
This means all probes in a single iteration $j$ can be simulated with no more number of shares of $(\text{\pmb{T}}_i)$.
It is clear that if an adversary can place $t$ probes across different iterations, these can also be simulated with no more number of input shares if $\mathbf{c}$ is given to the simulator.
As a result, the entire gadget \texttt{SecRowEch} is $t$-NIo secure when $\mathbf{c}$ is public.
\qed

\subsection{Masked Back Substitution with Public Output}\label{sec:SecBackSub}
The second and final step in solving a system of linear equations ($\text{\pmb{A}}\mathbf{x}=\mathbf{b}$), is performing back substitution on a system (e.g., matrix) in row-echelon form ($[\text{\pmb{A}}_i ~|~ \mathbf{b}_i]$) and is Boolean shared.
We propose to compute the unique solution $\mathbf{x}$ as a public output, with a secure recombination of the shares.

\begin{algorithm}
\DontPrintSemicolon
\KwData{1. A Boolean sharing $(\text{\pmb{A}}_i)$ of matrix $\text{\pmb{A}} \in \mathbb{F}_q^{m \times m}$\\
\quad \quad \quad 
2. A Boolean sharing $(\mathbf{b}_i)$ of the vector $\mathbf{b} \in \mathbb{F}_q^m$.}
\KwResult{Unique, public solution $\mathbf{x} \in \mathbb{F}_q^m$ such that $\text{\pmb{A}} \mathbf{x} = \mathbf{b}$}
\BlankLine
\For{$j = m$ downto $2$}{
  $\mathbf{x}{\scriptstyle [j]} = \mathtt{FullAdd}((\mathbf{b}{\scriptstyle [j]}_i))$\;
  \For{$k = 1$ upto $j-1$}{
    $(\mathbf{b}{\scriptstyle [k]}_i) := (\mathbf{b}{\scriptstyle [k]}_i + \mathbf{x}{\scriptstyle [j]} \cdot \text{\pmb{A}}{\scriptstyle [k,j]}_i)$\;
  }
}
$\mathbf{x}{\scriptstyle [1]} = \mathtt{FullAdd}((\mathbf{b}{\scriptstyle [1]}_i))$\;
\Return{$\mathbf{x}$}
\caption{\texttt{SecBackSub}}\label{alg:backsub}
\end{algorithm}
More precisely, the system is solved by moving from the final row ($m$) to the first one, as in typical back substitution.
Firstly, the solution of the current row $\mathbf{x}{\scriptstyle [j]} = (\mathbf{b}{\scriptstyle [j]}_i)$ is securely unmasked.
Secondly, all the elements above in column $j$ of $(\text{\pmb{A}}_i)$ are made zero, by (securely) multiplying all its elements $(\text{\pmb{A}}{\scriptstyle [1:j-1,j]}_i)$ with the solution $\mathbf{x}{\scriptstyle [j]}$.
As the multiplier is unmasked, the operation can be performed share-wise.
For each row, that result is subtracted from the element in vector $\mathbf{b}$, share by share. 
Finally, this process is repeated for all rows except the first one, which can be directly solved and unmasked.

\subsubsection{Complexity}\hfill\\
% \textcolor{red}{TODO, based on complexity of individual gadgets.}
The run-time and randomness complexity of \texttt{SecBackSub} are: 
\begin{align*}
    T_{\mathtt{SecBackSub}}(n,m) &= (m-1)\cdot T_{\mathtt{FullAdd}}(n) + (\frac{m(m-1)}{2}\cdot 2n) + T_{\mathtt{FullAdd}}(n)\\
    %&= m\cdot(T_{\mathtt{RefreshMasks}}(n)+(n-1))+(\frac{m(m-1)}{2}\cdot 2n)\\
    %&= m\cdot(\frac{3n^2-3n}{2}+(n-1))+m(m-1)\cdot n \\
    &= \frac{3}{2}n^2m-\frac{3}{2}mn-m+m^2n\,,\\
% \end{align*}
% \begin{align*}
    R_{\mathtt{SecBackSub}}(n,m,w) &= (m-1)\cdot R_{\mathtt{FullAdd}}(n,w) + 0 + R_{\mathtt{FullAdd}}(n,w)\\
    %&= m\cdot R_{\mathtt{RefreshMasks}}(n,w) = m\cdot \frac{n(n-1)}{2}\cdot w 
    &= \frac{(n^2-n)mw}{2}\,.
\end{align*}
\subsubsection{Security}\hfill\\
We now argue about the first- and high-order security of Algorithm~\ref{alg:backsub} by proving it to be $t$-NIo secure with $n=t+1$ shares and public output $\mathbf{x}$. 
This means it provides resistance against an adversary with $t$ probes and allows the algorithm to be used in larger compositions.
\begin{lemma}\label{lem:SecBackSub}
The gadget $\mathtt{SecBackSub}$ (Algorithm~\ref{alg:backsub}) is $t$-NIo secure with public output $\mathbf{x}$.
\end{lemma}
\noindent\textit{Proof.} 
We first show that a single iteration $j$ is $t$-NIo secure with output $\mathbf{x}{\scriptstyle [j]}$, of which an abstract diagram is shown in Figure \ref{fig:backsub}.
We model the extraction of element $j$ and column $j$ from vector $(\mathbf{b}_i)$ and matrix $(\text{\pmb{A}}_i)$ as $t$-NI gadgets $G_1$-$G_2$ and $G_3$, respectively.
We also model the loop of share-wise multiplications and additions in Line 3-4 as a single $t$-NI gadget, which can be trivially shown as the operations are performed share by share and each iteration is independent.
As a result, the iterations are assumed to be executed in parallel, and we summarize them into a single gadget $G_5$.
An adversary can probe the intermediate values $t_{G_i}$ and output shares $o_{G_i}$ of each gadget $G_i$, except the outputs of the entire gadget.
The total number of probes in Algorithm \ref{alg:backsub} is defined as:
\begin{equation*}
    t_{A_{\text{\ref{alg:backsub}}}} = \sum\limits_{i=1}^{5}t_{G_i} + \sum\limits_{i=1}^{4}o_{G_i}
\end{equation*}
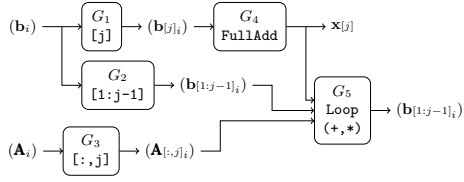
\begin{figure}[t]
    \centering
    % \begin{frame}{}
        % \centering
        \scalebox{0.65}{\setlength{\nodedistance}{4mm}
\begin{tikzpicture}[ % <--- used style are moved here 
node distance = \nodedistance,
data/.style = {},
NIop/.style = {rounded corners, draw=black},
SNIop/.style = {double, double distance=0.5mm, rounded corners, draw=black}]

% INPUTS
\node (I_b) [data]  {$(\mathbf{b}_i)$};
\node (I_A) [data, below=5\nodedistance of I_b]  {$(\text{\pmb{A}}_i)$};

\coordinate [right=\nodedistance of I_b] (tmp1);
\coordinate [below=3\nodedistance of tmp1] (tmp2);

% GADGETS
\node (G1_jselect) [NIop, right=\nodedistance of tmp1] {\begin{tabular}{c} $G_1$ \\ \texttt{[j]}\end{tabular}};

\node (G2_kselect) [NIop, right=\nodedistance of tmp2] {\begin{tabular}{c} $G_2$ \\ \texttt{[1:j-1]}\end{tabular}};

\node (G3_jselect) [NIop, right=\nodedistance of I_A] {\begin{tabular}{c} $G_3$ \\ \texttt{[:,j]}\end{tabular}};

\node (bj) [data, right=\nodedistance of G1_jselect]  {$(\mathbf{b}{\scriptstyle [j]}_i)$};

\node (bk) [data, right=\nodedistance of G2_kselect]  {$(\mathbf{b}{\scriptstyle [1:j-1]}_i)$};

\node (Aj) [data, right=\nodedistance of G3_jselect]  {$(\text{\pmb{A}}{\scriptstyle [:,j]}_i)$};

\node (G4_fulladd) [NIop, right=\nodedistance of bj] {\begin{tabular}{c} $G_4$ \\ \texttt{FullAdd}\end{tabular}};

\coordinate[right=\nodedistance of G4_fulladd] (tmp3);
\coordinate[right=\nodedistance of bk] (tmp4);
\coordinate[right=\nodedistance of Aj] (tmp5);

\node (xj) [data, right=\nodedistance of tmp3] {$\mathbf{x}{\scriptstyle [j]}$};

\node (G5_loop) [NIop, below=2\nodedistance of xj] {\begin{tabular}{c} $G_5$ \\ \texttt{Loop} \\ \texttt{(+,*)}\end{tabular}};

% OUTPUTS
\node (O_b) [data, right=\nodedistance of G5_loop]  {$(\mathbf{b}{\scriptstyle [1:j-1]}_i)$};

% CONNECT
\draw [->] (I_b) -- (G1_jselect);
\draw[->] (I_A) -- (G3_jselect);
\draw[->] (G3_jselect) -- (Aj);
\draw[->] (G1_jselect) -- (bj);
\draw[->] (tmp1) -- (tmp2) -- (G2_kselect);
\draw[->] (G2_kselect) -- (bk);
\draw[->] (bj) -- (G4_fulladd);
\draw[->] (G4_fulladd) -- (xj);
\draw[->] (tmp3) -- (tmp3 |- G5_loop.160) -- (G5_loop.160);
\draw[->] (bk) -- (tmp4) -- (tmp4 |- G5_loop) -- (G5_loop);
\draw[->] (Aj) -- (tmp5) -- (tmp5 |- G5_loop.200) -- (G5_loop.200);
\draw[->] (G5_loop) -- (O_b);

\end{tikzpicture}}
    % \end{frame}
    \caption{An abstract diagram of a single iteration $j$ of \texttt{SecBackSub} (Algorithm \ref{alg:backsub}). 
    The $t$-NI gadgets are depicted with a single border, the $t$-SNI gadgets with a double border.}
    \label{fig:backsub}
% \vspace{-10pt}
\end{figure}
To prove Lemma \ref{lem:SecBackSub}, we show that the internal and output probes of each gadget in Algorithm \ref{alg:backsub} can be perfectly simulated with $\leq t_{A_{\text{\ref{alg:backsub}}}}$ input shares.
As $t$-NI security implies that the simulation of internal and output probes of a gadget $G_i$ requires a corresponding number of shares of its input.
From this, it is clear that on a matrix/vector-level, the probes cannot be perfectly simulated as there are duplicate entries in the sum ($2 \cdot t_{G_5}$ of $(\mathbf{b}_i)$).
To overcome this issue, we model gadget $G_5$ to operate on individual entries in row/column vectors rather than on the complete variables.
As a result, to simulate $t_{G_5}$ intermediate values requires $t_{G_5}$ shares of element $j$ and elements $[1:j-1]$.
As the coefficients inside a column/row are independent, the simulation succeeds.
Following the flow from the output to the inputs, all probes required for simulation are summed up. 
We determine the set of shares of the inputs $|I|$, required for simulating the entire algorithm.
As $|I| = t_{G_1} + o_{G_1} + t_{G_2} + o_{G_2} + t_{G_3} + o_{G_3} + t_{G_4} + o_{G_4} + t_{G_5} \leq t_{A_{\text{\ref{alg:backsub}}}}$, the iteration $j$ is $t$-NI.
We have shown that all $t$ probes placed by an adversary in a single iteration $j$ can be simulated with no more number of shares of inputs $(\text{\pmb{A}}_i)$ and $(\mathbf{b}_i)$.
As a result, all probes across different iterations can also be simulated with no more number of input shares.
\qed

% \subsection{Back Substitution with Masked Output (?)}

% \subsection{Overall security and Complexity}\label{sec:security}

% \begin{table}[h]
% \centering
% \begin{tabular}{ lcc } 
%  \hline
%  {\bf Gadget} & {\bf Required random bits per call } & {\bf Number of calls}  \\
%  \hline \\
%  $\mathtt{Refresh}$     & $m (n - m) \log q$         & $1$ \\
%  $\mathtt{SecMatVec}$   & $m \log q$                 & $m^2$ \\
%  $\mathtt{FullXor}$     & $m^2 (m + 1) \log q / 2$   & $1$ \\
%  \hline
% \end{tabular}
% \caption{Required random bits in $\mathtt{MaskedCompactKeyGen}$}
% \label{tab:keygencomp}
% \end{table}

% Table \ref{tab:keygencomp} shows $\mathtt{MaskedCompactKeyGen}$ requires $\mathcal{O} (m^3 \log q)$ random bits, apart from randomness used by $\mathtt{MaskedExpand}_{sk}$ gadget.

% \begin{table}[h]
% \centering
% \begin{tabular}{ lcc } 
%  \hline
%  {\bf Gadget} & {\bf Required random bits per call } & {\bf Number of calls}  \\
%  \hline \\
%  $\mathtt{SecMatVec}$   & $m \log q$                 & $\mathcal{O}(m)$ \\
%  $\mathtt{SecQuad}$     & $n \log q$                 & $\mathcal{O}(1)$ \\
%  $\mathtt{RowEchelon}$  & $\mathcal{O}(m^3 \log q)$  & $\mathcal{O}(1)$ \\
%  $\mathtt{BackSub}$     & $m \log q$                 & $1$ \\
%  $\mathtt{FullXor}$     & $(n - m) \log q$           & $1$ \\
%  \hline
% \end{tabular}
% \caption{Required random bits in $\mathtt{MaskedSign}$}
% \label{tab:signcomp}
% \end{table}

% Table \ref{tab:signcomp} shows gadget $\mathtt{MaskedSign}$ requires $\mathcal{O} (m^3 \log q)$ random bits, apart from randomness used by $\mathtt{MaskedExpand}_{\pmb{v}}$ gadget.

\FloatBarrier

\section{Application to Different Digital Signature Schemes}\label{sec:costcomparison}

\begin{table}[t]
\vspace{-15pt}
\scriptsize
\centering
\caption{Cost estimation of masked Gaussian elimination to different digital signature schemes. Here, $m$ represents the dimension of the matrix $\text{\pmb{A}}$, $q$ represents the modulus, and $n$ represents the number of masking shares.}
\label{tab:costcomparison}
\begin{tabular}{l|c|rr|rrr|rrr}
\hline
\multicolumn{1}{c|}{{\color[HTML]{3531FF} }} & {\color[HTML]{3531FF} } & \multicolumn{2}{c|}{{\color[HTML]{3531FF} \textbf{Scheme}}} & \multicolumn{3}{c|}{{\color[HTML]{3531FF} \textbf{Operations}}} & \multicolumn{3}{c}{{\color[HTML]{3531FF} \textbf{Randomness}}} \\
\multicolumn{1}{c|}{{\color[HTML]{3531FF} }} & {\color[HTML]{3531FF} } & \multicolumn{2}{c|}{{\color[HTML]{3531FF} \textbf{Parameters}}} & \multicolumn{3}{c|}{{\color[HTML]{3531FF} \textbf{(x1000)}}} & \multicolumn{3}{c}{{\color[HTML]{3531FF} \textbf{(KB)}}} \\ \cline{3-10} 
\multicolumn{1}{c|}{\multirow{-3}{*}{{\color[HTML]{3531FF} \textbf{\begin{tabular}[c]{@{}c@{}}Scheme\end{tabular}}}}} & \multirow{-3}{*}{{\color[HTML]{3531FF} \textbf{\begin{tabular}[c]{@{}c@{}}NIST\\ Security\\ Level\end{tabular}}}} & {\color[HTML]{3531FF} \textbf{$q$}} & {\color[HTML]{3531FF} \textbf{$m$}} & {\color[HTML]{3531FF} \textbf{$n=2$}} & {\color[HTML]{3531FF} \textbf{$n=3$}} & {\color[HTML]{3531FF} \textbf{$n=4$}} & {\color[HTML]{3531FF} \textbf{$n=2$}} & {\color[HTML]{3531FF} \textbf{$n=3$}} & {\color[HTML]{3531FF} \textbf{$n=4$}} \\ \hline
{\color[HTML]{036400} } &  & 256 & 44 & 105 & 260 & 482 & 742 & 2226 & 4452 \\
{\color[HTML]{036400} } & \multirow{-2}{*}{I} & 16 & 64 & 300 & 747 & 1392 & 1112 & 3336 & 6671 \\
{\color[HTML]{036400} } & III & 256 & 72 & 428 & 1065 & 1986 & 3146 & 9437 & 18873 \\
\multirow{-4}{*}{{\color[HTML]{036400} \begin{tabular}[c]{@{}l@{}}UOV\\ ~\cite{uov_spec}\end{tabular}}} & V & 256 & 96 & 985 & 2459 & 4590 & 7360 & 22079 & 44158 \\ \hline
{\color[HTML]{036400} } & I & 16 & 64 & 300 & 747 & 1392 & 1112 & 3336 & 6671 \\
{\color[HTML]{036400} } & III & 16 & 96 & 973 & 2434 & 4546 & 3680 & 11040 & 22079 \\
\multirow{-3}{*}{{\color[HTML]{036400} \begin{tabular}[c]{@{}l@{}}MAYO\\ ~\cite{mayo_spec}\end{tabular}}} & V & 16 & 128 & 2263 & 5670 & 10597 & 8638 & 25914 & 51827 \\ \hline
{\color[HTML]{036400} } &  & 7 & 100 & 1084 & 2717 & 5076 & 3102 & 9306 & 18612 \\
{\color[HTML]{036400} } &  & 31 & 60 & 249 & 619 & 1155 & 1147 & 3441 & 6881 \\
{\color[HTML]{036400} } &  & 31 & 70 & 388 & 968 & 1806 & 1806 & 5417 & 10834 \\
{\color[HTML]{036400} } & \multirow{-4}{*}{I} & 127 & 54 & 184 & 457 & 852 & 1175 & 3524 & 7048 \\
{\color[HTML]{036400} } &  & 7 & 140 & 2922 & 7333 & 13712 & 8431 & 25292 & 50584 \\
{\color[HTML]{036400} } &  & 31 & 87 & 730 & 1825 & 3407 & 3432 & 10295 & 20590 \\
{\color[HTML]{036400} } &  & 31 & 100 & 1097 & 2744 & 5125 & 5184 & 15550 & 31100 \\
{\color[HTML]{036400} } & \multirow{-4}{*}{III} & 127 & 78 & 531 & 1327 & 2476 & 3472 & 10415 & 20829 \\
{\color[HTML]{036400} } &  & 7 & 190 & 7217 & 18131 & 33918 & 20942 & 62825 & 125650 \\
{\color[HTML]{036400} } &  & 31 & 114 & 1610 & 4032 & 7533 & 7646 & 22937 & 45873 \\
{\color[HTML]{036400} } &  & 31 & 120 & 1872 & 4689 & 8761 & 8904 & 26710 & 53419 \\
\multirow{-12}{*}{{\color[HTML]{036400} \begin{tabular}[c]{@{}l@{}}QR-UOV\\ ~\cite{furue2023qr}\end{tabular}}} & \multirow{-4}{*}{V} & 127 & 105 & 1265 & 3166 & 5914 & 8373 & 25119 & 50237 \\ \hline
{\color[HTML]{036400} } &  & 16 & 68 & 357 & 890 & 1660 & 1329 & 3987 & 7974 \\
{\color[HTML]{036400} } &  & 16 & 72 & 421 & 1051 & 1961 & 1573 & 4719 & 9437 \\
{\color[HTML]{036400} } & \multirow{-3}{*}{I} & 16 & 80 & 572 & 1428 & 2666 & 2147 & 6439 & 12877 \\
{\color[HTML]{036400} } &  & 16 & 100 & 1097 & 2744 & 5125 & 2147 & 6439 & 12877 \\
{\color[HTML]{036400} } &  & 16 & 99 & 1065 & 2664 & 4976 & 4031 & 12093 & 24186 \\
{\color[HTML]{036400} } & \multirow{-3}{*}{III} & 16 & 128 & 2263 & 5670 & 10597 & 8638 & 25914 & 51827 \\
{\color[HTML]{036400} } &  & 16 & 132 & 2477 & 6209 & 11604 & 9465 & 28395 & 56789 \\
{\color[HTML]{036400} } &  & 16 & 135 & 2647 & 6634 & 12400 & 10119 & 30356 & 60712 \\
\multirow{-9}{*}{{\color[HTML]{036400} \begin{tabular}[c]{@{}l@{}}SNOVA\\ ~\cite{snova_spec}\end{tabular}}} & \multirow{-3}{*}{V} & 16 & 160 & 4369 & 10959 & 20489 & 16773 & 50317 & 100634 \\ \hline
{\color[HTML]{036400} } & I & 256 & 46 & 119 & 294 & 547 & 845 & 2534 & 5068 \\
{\color[HTML]{036400} } & III & 256 & 72 & 428 & 1065 & 1986 & 3146 & 9437 & 18873 \\
\multirow{-3}{*}{{\color[HTML]{036400} \begin{tabular}[c]{@{}l@{}}MQ-Sign\\ ~\cite{mq_sign_spec}\end{tabular}}} & V & 256 & 96 & 985 & 2459 & 4590 & 7360 & 22079 & 44158 \\ \hline
\end{tabular}
\vspace{-10pt}
\end{table}

This section discusses the operational and randomness cost of the masked Gaussian elimination on multivariate- and code-based digital signatures. 
In total, we observed that at least the following ten UOV-based digital signatures use some variant of Gaussian elimination: UOV, MAYO, SNOVA, QR-UOV, MQ-Sign, PROV, VOX, TUOV, VDOO, IPRainbow, and WAVE. 
In between these schemes, UOV, MAYO, SNOVA, and QR-UOV have advanced to the second round of the NIST additional digital signature standardization procedure~\cite{NISTDS_R2}, and MQ-Sign is a second-round candidate for the Korean PQC standardization procedure~\cite{KPQC_additional_sign_round2}. 
Therefore, we have restricted our cost analysis to these five multivariate-based digital signature candidates in this work. 
As seen from the complexity analysis in Section~\ref{sec:maskinguov}, the cost of the masked Gaussian elimination mainly depends on the dimension of the matrix $\text{\pmb{A}}$ ($m$), the width of the modulus $q$ ($w$), and the number of shares ($n$). 

% \textcolor{black}{Wave is a code-based signature that follows the hash-and-sign paradigm. It was also submitted to the NIST additional round DS standardization process. The Gaussian elimination is used in the key-generation and signature phase of the Wave signature, in particular the decoding phase. For more details, we refer to the specification document of Wave signature. So similarly, we can use our algorithm (or a slight modification) to protect Gaussian elimination.}
We present a cost analysis of the five selected schemes in Table~\ref{tab:costcomparison}. 
\textcolor{black}{We have used the parameters of the corresponding scheme in the run-time and randomness complexity equations to calculate the expected arithmetic operational and randomness cost of the masked Gaussian elimination for each scheme, respectively.}
It can be seen from the table that the effect of the matrix dimension $m$ and the share count $n$ is much more potent in operation and randomness costs than the width of the modulus $q$. 
For instance, in the two versions of UOV for the NIST security level I, the operational cost of masked GE in UOV-Ip ($m=44\ \&\ q=256$) is $2.9\times$ less compared to UOV-Is ($m=64\ \&\ q=16$) for first-, second-, and third-order masking. The randomness cost of GE in UOV-Ip is $1.5\times$ less compared to UOV-Is for first, second, and third-order masking. 
% Due to similar reasons, the operational and randomness costs of masked Gaussian elimination in UOV are lower than those of MAYO for second and third-order masking. 
Due to similar reasons, the operational cost of first-, second-, and third-order masked GE is 2.3$\times$ higher, and the randomness cost is 1.2$\times$ higher in MAYO compared to UOV for III and V, both security levels. 
% Due to similar reasons, the operational and randomness costs of first-, second-, and third-order masked GE are 2.3$\times$ and 1.2$\times$ higher in MAYO compared to UOV for III and V, both security levels. 
% For security level I, the operational and randomness costs of masked GE are 2.9$\times$ and 1.5$\times$ higher for the MAYO (similar to UOV-Is) than UOV-Ip.

As the modulus $q$ in QR-UOV is prime, we considered the next closest power-of-two integer as the modulus during the complexity cost analysis of masked Gaussian elimination on QR-UOV. 
However, the actual cost is likely to be higher because a masked prime modulus reduction is often more expensive than a masked power-of-two modulus reduction. 
We can also observe from Table~\ref{tab:costcomparison} that the operational and randomness costs of masked GE in QR-UOV and SNOVA even vary hugely for the same security level. 
In fact, two NIST security level III variants ((i) $m=87\ \&\ q=31$, and (ii) $m=78\ \&\ q=127$) of QR-UOV are cheaper to protect than a NIST security level I variant ($m=100\ \&\ q=7$) of QR-UOV. 
We conclude that the main contributors to operational and randomness cost in masked GE with back substitution are matrix dimension $m$ and masking order $t=n-1$.

% \textcolor{black}{The code-based signature scheme, Wave, follows the hash-and-sign paradigm and uses Gaussian elimination in the signature algorithm, particularly in the decoding phase. So, we can use our masked algorithms (with slight modifications) to protect the Gaussian elimination used in WAVE.}

% \subsection{Estimation}

\section{Implementation Results}\label{sec:performance}
In this section, we discuss our software (C, M4) implementation of the gadgets introduced in Section \ref{sec:maskinguov} and evaluate their performance for different UOV parameter sets and sharing degrees.
The performance results are obtained from running our code\footnote{arm-none-eabi-gcc v10.3.1}, which we make publicly available, on the STM32Discovery board, containing an STM32 Arm Cortex-M4 microcontroller.
We use the on-chip TRNG for on-the-fly randomness generation, which is included in the total cycle count.
Below, we discuss and show results for practically relevant security orders ($n=2,3,4$), but our implementation could be scaled up to arbitrary order.
We emphasize that our implementation is only a proof-of-concept.
Mitigation of micro-architectural leakages and optimized masked implementations are beyond the scope of this work and left as future work.

\begin{table}[tbh]
\scriptsize
\centering
\caption{Detailed performance overview (cycle counts $\times 1000$) of masked Gaussian Elimination ($n=2,3,4$) for UOV-I, -III, -V on Arm Cortex-M4.}
\label{tab:UOVperf}
\scalebox{0.9}{
\begin{tabular}{l|lll|r|lrllrllrl}
\hline
{\color[HTML]{3531FF} } & \multicolumn{3}{c|}{{\color[HTML]{3531FF} }} & \multicolumn{1}{c|}{{\color[HTML]{3531FF} }} & \multicolumn{9}{c}{{\color[HTML]{3531FF} \textbf{Masked}}} \\ \cline{6-14} 
\multirow{-2}{*}{{\color[HTML]{3531FF} \textbf{Scheme}}} & \multicolumn{3}{c|}{\multirow{-2}{*}{{\color[HTML]{3531FF} \textbf{Operation}}}} & \multicolumn{1}{c|}{\multirow{-2}{*}{{\color[HTML]{3531FF} \textbf{Umasked}}}} &  & \multicolumn{2}{c|}{{\color[HTML]{3531FF} \textbf{1st-order}}} &  & \multicolumn{2}{c|}{{\color[HTML]{3531FF} \textbf{2nd-order}}} &  & \multicolumn{2}{c}{{\color[HTML]{3531FF} \textbf{3rd-order}}} \\ \hline
{\color[HTML]{3531FF} } & \multicolumn{3}{l|}{{\color[HTML]{036400} GE (Total)}} & 1034 &  & \textbf{15649} & \multicolumn{1}{l|}{\textbf{(15.1$\times$)}} &  & \textbf{39587} & \multicolumn{1}{l|}{\textbf{(38.3$\times$)}} &  & \textbf{69186} & \textbf{(66.9$\times$)} \\
{\color[HTML]{3531FF} } & \multicolumn{1}{l|}{{\color[HTML]{036400} }} & \multicolumn{2}{l|}{{\color[HTML]{036400} SecRowEch}} & 999 &  & 15575 & \multicolumn{1}{l|}{(15.6$\times$)} &  & 39469 & \multicolumn{1}{l|}{(39.5$\times$)} &  & 69021 & (69.1$\times$) \\
{\color[HTML]{3531FF} } & \multicolumn{1}{l|}{{\color[HTML]{036400} }} & \multicolumn{1}{l|}{{\color[HTML]{036400} }} & {\color[HTML]{036400} Step 1} & 80 &  & 1998 & \multicolumn{1}{l|}{(25$\times$)} &  & 5642 & \multicolumn{1}{l|}{(70.5$\times$)} &  & 9901 & (123.8$\times$) \\
{\color[HTML]{3531FF} } & \multicolumn{1}{l|}{{\color[HTML]{036400} }} & \multicolumn{1}{l|}{{\color[HTML]{036400} }} & {\color[HTML]{036400} Step 2} &  &  & 39 & \multicolumn{1}{l|}{} &  & 108 & \multicolumn{1}{l|}{} &  & 193 &  \\
{\color[HTML]{3531FF} } & \multicolumn{1}{l|}{{\color[HTML]{036400} }} & \multicolumn{1}{l|}{{\color[HTML]{036400} }} & {\color[HTML]{036400} Step 3} &  &  & 598 & \multicolumn{1}{l|}{} &  & 1335 & \multicolumn{1}{l|}{} &  & 2426 &  \\
{\color[HTML]{3531FF} } & \multicolumn{1}{l|}{{\color[HTML]{036400} }} & \multicolumn{1}{l|}{{\color[HTML]{036400} }} & {\color[HTML]{036400} Step 4} &  &  & 12938 & \multicolumn{1}{l|}{} &  & 32383 & \multicolumn{1}{l|}{} &  & 56498 &  \\
\multirow{-7}{*}{{\color[HTML]{3531FF} \begin{tabular}[c]{@{}l@{}}UOV-I\\ \\ (m=44 \\ q=256)\end{tabular}}}  & \multicolumn{1}{l|}{{\color[HTML]{036400} }} & \multicolumn{2}{l|}{{\color[HTML]{036400} SecBackSub}} & 32 &  & 74 & \multicolumn{1}{l|}{(2.3$\times$)} &  & 118 & \multicolumn{1}{l|}{(3.7$\times$)} &  & 164 & (5.1$\times$) \\\hline
%& \multicolumn{1}{l|}{{\color[HTML]{036400} }} & \multicolumn{2}{l|}{{\color[HTML]{036400} Others}} & 3 &  & - & \multicolumn{1}{l|}{} &  & 11 & \multicolumn{1}{l|}{(1.8$\times$)} &  & 17 & (1.9$\times$) \\ \hline
{\color[HTML]{3531FF} UOV-III} & \multicolumn{3}{l|}{{\color[HTML]{036400} GE (Total)}} & 4117 &  & \textbf{62680} & \multicolumn{1}{l|}{\textbf{(15.2$\times$)}} &  & \textbf{157901} & \multicolumn{1}{l|}{\textbf{(38.4$\times$)}} &  & \textbf{275768} & \textbf{(67$\times$)} \\ \hline
{\color[HTML]{3531FF} UOV-V} & \multicolumn{3}{l|}{{\color[HTML]{036400} GE (Total)}} & 9336 &  & \textbf{143361} & \multicolumn{1}{l|}{\textbf{(15.4$\times$)}} &  & \textbf{360590} & \multicolumn{1}{l|}{\textbf{(38.6$\times$)}} &  & \textbf{1025942} & \textbf{(109.9$\times$)} \\ \hline
\end{tabular}
}
\end{table}

\subsection{Case Study: UOV}
%discussion table 
% masking factor
Table~\ref{tab:UOVperf} shows the performance results (clock cycles) of our masked implementation for all three UOV security levels and scaled to first-, second- and third-masking order. We include the performance numbers of the unmasked UOV reference code \cite{uov_spec}, of the supported parameter sets.
We repeat that the randomness generation overhead is included in the benchmarking results. %, and total numbers include data alignment (`Other').
The full masked GE operation results in about 15.1/15.2/15.4$\times$ overhead for 2 shares at NIST security level I/III/V. 
As expected, the performance overhead increases significantly for higher-order protected implementations. Our second-order masked implementation has overhead factor 38.3/38.4/38.6$\times$ for UOV-I, -III and -V, respectively. The third-order implementation has an overhead of 66.9-109.9$\times$ for different NIST security levels.
The main contributor to the cycle count is the conversion to row-echelon form, more specifically, Steps 1 and 4 ($\sim 95$\%). 
At all protection orders, the repeated execution of \texttt{SecMultSub} and \texttt{SecCondAdd} are the main bottlenecks and good targets for future optimizations.

%optimization UOV spec
We observe that the masked implementation of Step 1, which makes the pivot element non-zero, has an overhead of $\sim 25/70/124\times$ for 2/3/4 shares.
The authors of UOV propose an optimization for the iteration of conditional additions (Algorithm~\ref{alg:ge}), which we also implemented.
Instead of iterating over all rows below the current pivot row (up to $m$), the UOV submission describes iterating over (and conditionally adding) only a few rows below the pivot row.
The exact amount depends on the parameter selection but ensures a sufficiently low probability of the pivot element being non-zero.
We encourage other UOV-like candidates to explore similar optimizations, which will lead to more efficient (masked) implementations of GE with back substitution.

\section{Conclusions}\label{sec:conclusion}
In this work, we presented first- and higher-order masked algorithms for Gaussian elimination with back substitution: \texttt{SecRowEch} \& \texttt{SecBackSub}. 
We analyze several novel multivariate- and code-based PQC schemes and show that GE is a critical operation for solving a system of linear equations and requires side-channel protection.
Our \texttt{SecCondAdd} gadget allows us to make a pivot element nonzero by conditionally adding other rows in the matrix without revealing if it is zero.
We rely on the \texttt{SecScalarMult} gadget to efficiently multiply a matrix row with the (masked) inverse of its pivot element to make the pivot element one.
Our approach only requires a single mask conversion.
For the same masking order, the matrix dimension $m$ is the main contributor to operation and randomness cost in masked GE. 
The MAYO scheme has a larger $m$ compared to UOV, and as a result, the GE is 2.3$\times$ and 1.2$\times$ more expensive to mask.
Future work includes analysis of reduced iteration counts in the GE, as proposed in the UOV specification, for other PQC DS candidates.
We implement our algorithms in C for arbitrary protection orders and parameter sets of different PQC schemes. 
We also evaluate its performance on Arm Cortex-M4 platforms.
In future work, a complete masking and hardening against physical attacks for all mentioned schemes can be constructed using the methods proposed in this work.

\noindent\textbf{Acknowledgements.} This work was partially supported by Horizon 2020 ERC Advanced Grant (101020005 Belfort), Horizon Europe (101070008 ORSHIN), CyberSecurity Research Flanders with reference number VOEWICS02, BE QCI: Belgian-QCI (3E230370) (see beqci.eu), and Intel Corporation. Anindya Ganguly is supported by TCS research fellowship. The work of Angshuman Karmakar is supported by the Research-I foundation from Infosys, the Initiation grant from IIT Kanpur, and the Google India research fellowship.
\FloatBarrier

\bibliographystyle{splncs04}
% \bibliography{mask,crypto/crypto}
\bibliography{crypto/abbrev3,crypto/crypto,main}

% \newpage
\appendix
\section{Auxiliary Algorithms}\label{sec:appen}
\begin{algorithm}[tbh]
\DontPrintSemicolon
\KwData{A Boolean sharing $(x_i)$ of $x \in \mathbb{F}_q$}
\KwResult{A Boolean sharing $(y_i)$ of $y \in \mathbb{F}_q$ such that $\sum\limits_{i=1}^ny_i = \sum\limits_{i=1}^nx_i$}
\BlankLine
$(y_i) := (x_i)$\;
\For{$i=2$ upto $n$}{
    $r \leftarrow \mathbb{F}_q$\;
    $y_1 = y_1 + r$\;
    $y_i = y_i - r$\;
}
\Return{$(y_i)$}
\caption{\texttt{Refresh}, from \cite{CCS:BFGGHS16}}\label{alg:refresh}
\end{algorithm}

\vspace{-15pt}
\begin{algorithm}[tbh]
\DontPrintSemicolon
\KwData{A Boolean sharing $(x_i)$ of $x \in \mathbb{F}_q$}
\KwResult{A Boolean sharing $(y_i)$ of $y \in \mathbb{F}_q$ such that $\sum\limits_{i=1}^ny_i = \sum\limits_{i=1}^nx_i$}
\BlankLine
$(y_i) := (x_i)$\;
\For{$i=1$ upto $n$}{
    \For{$j=i+1$ upto $n$}{
        $r \leftarrow \mathbb{F}_q$\;
        $y_i = y_i + r$\;
        $y_j = y_j - r$\;
    }
}
\Return{$(y_i)$}
\caption{\texttt{StrongRefresh}, from \cite{CCS:BFGGHS16}}\label{alg:strongrefresh}
\end{algorithm}

\vspace{-15pt}
\begin{algorithm}[tbh]
\DontPrintSemicolon
\KwData{A Boolean sharing $(y_i)$}
\KwResult{Unmasked value $y$ such that $y=\sum\limits_{i=1}^ny_i$}
\BlankLine
$(a_i) := \mathtt{StrongRefresh}((y_i))$\tcc*{free-$t$-SNI}
$y := a_1 + \cdots + a_n$\; %\tcc*{\cite{coron2014higher}}
\Return{$y$}
\caption{\texttt{FullAdd}, from \cite{coron2014secure,barthe2018masking}}\label{alg:fullxor}
\end{algorithm}

\vspace{-15pt}
\begin{algorithm}[tbh]
\DontPrintSemicolon
\KwData{1. A Boolean sharing $(x_i)$ of a coefficient $x \in \mathbb{F}_q$.\\
\quad \quad \quad
2. Parameter $w=\lceil \log (q) \rceil$}
\KwResult{One-bit Boolean sharing $(b_i)$ such that $\sum\limits_{i=1}^nb_i = 0 \Leftrightarrow \sum\limits_{i=1}^n x_i = 0$}
\BlankLine
$(t_i) := (x_i)$\;
$\text{len} := w/2$\;
\While{$\text{len} \geq 1$}{
    $(l_i) := \mathtt{StrongRefresh}((t_i^{[2\text{len}:\text{len}]}))$\;
    $(r_i) := (t_i^{[\text{len}:1]})$\;
    $(t_i) = \mathtt{SecOR}((l_i), (r_i))$ \tcc*{\cite{chen2024masking}}
    $\text{len} = \text{len} \gg 1$
}
\Return{$(t_i^{[1]})$}
\caption{\texttt{SecNonzero}, from \cite{chen2024masking}}\label{alg:secnonzero}
\end{algorithm}

\vspace{-15pt}
\begin{algorithm}[tbh]
\DontPrintSemicolon
\KwData{A Boolean sharing $(x_i)$ of a coefficient $x \in \mathbb{F}_q$}
\KwResult{A multiplicative sharing $(m_i)$ such that $\sum\limits_{i=1}^nx_i = \prod\limits_{i=1}^n m_i$}
\BlankLine
$m_1 := x_1$\;
\For{$j=2$ upto $n$}{
    $m_j \leftarrow \mathbb{F}_q$\;
    $m_1 = m_1 * m_j$\;
    \For{$k=2$ upto $n-j+1$}{
        $r \leftarrow \mathbb{F}_q$\;
        $x_k = m_j * x_k$\;
        \texttt{\#\# Refresh additive share}\;
        $x_k = x_k + r$\;
        $m_1 = m_1 + x_k$\;
        $x_k = r$\;
    }
    $x_{n-j+2} = x_{n-j+2} * m_j$\;
    $m_1 = m_1 + x_{n-j+2}$\;
    $m_j = m_j^{-1}$
}
\Return{$(m_i)$}
\caption{\texttt{B2M}, from \cite{CHES:GenProQui11}}\label{alg:B2M}
\end{algorithm}

\clearpage
\FloatBarrier
\section{Complexity Analysis of \texttt{SecNonzero}}\label{sec:appen2}
% \textcolor{red}{TODO, based on complexity of individual gadgets.}
The run-time and randomness complexity of \texttt{SecNonzero} (Alg. \ref{alg:secnonzero}) are: 

\resizebox{.95\linewidth}{!}{
  \begin{minipage}{\linewidth}
\begin{align*}
    T_{\mathtt{SecNonzero}}(n) &= n + 1 + w\cdot (T_{\mathtt{StrongRefresh}}(n) + n + T_{\mathtt{SecOR}}(n) + 1) \\
    &= n + 1 + (\lceil\log (w+1)\rceil-1)\cdot(\frac{3n^2-3n}{2} + n + (2n+T_{\mathtt{SecAnd}}(n)+1) + 1)\\
    &= n + 1 + (\lceil\log (w+1)\rceil-1)\cdot(\frac{3n^2-3n}{2} + n + (2n + \frac{7n^2-5n}{2} + 1) + 1)\\
    &= (5n^2 + 2n - 1)+\lceil\log (w+1)\rceil\cdot(5n^2-n+2)\,,
\end{align*}
\end{minipage}
}
\begin{align*}
    R_{\mathtt{SecNonzero}}(n,w) &= \sum_{len=1}^{(\lceil\log (w+1)\rceil-1)}\cdot(R_{\mathtt{StrongRefresh}}(n,len) + R_{\mathtt{SecOR}}(n,len))\\
    &=\sum_{len=1}^{(\lceil\log (w+1)\rceil-1)}\cdot((\frac{n^2-n}{2}\cdot len) + R_{\mathtt{SecAnd}}(n,len))\\
    &=\sum_{len=1}^{(\lceil\log (w+1)\rceil-1)}\cdot(\frac{n^2-n}{2} + \frac{n^2-n}{2})\cdot len\\
    &= \frac{\lceil\log (w+1)\rceil^2-\lceil\log (w+1)\rceil}{2}\cdot(n^2-n)\,.
\end{align*}

\section{Complexity Analysis of \texttt{SecRowEch}}\label{sec:appen3}
The run-time and randomness complexity of \texttt{SecRowEch} (Alg. \ref{alg:rowech}) are: 

\resizebox{.9\linewidth}{!}{
  \begin{minipage}{\linewidth}
\begin{align*}
    T_{\mathtt{SecRowEch}}(n,m) =& \frac{m^2-m}{2}\cdot(T_{\mathtt{SecNonzero}}(n)+T_{\mathtt{SecNOT}}(n)) + \frac{2m^3+3m^2+m}{6}\cdot \\&T_{\mathtt{SecCondAdd}}(n,1) + m\cdot T_{\mathtt{SecNonzero}}(n) + m\cdot T_{\mathtt{FullAdd}}(n) + m \\&+ m\cdot T_{\mathtt{B2Minv}}(n) + \frac{m^2+3m}{2}\cdot T_{\mathtt{SecScalarMult}}(n,1) + \frac{m^2-m}{2}\cdot\\&T_{\mathtt{StrongRefresh}}(n) + \frac{2m^3+3m^2+m}{6}\cdot T_{\mathtt{SecMultAdd}}(n,1)\\
    =& \frac{m^2-m}{2}\cdot(((5n^2 + 2n - 1)+\lceil\log (w+1)\rceil\cdot(5n^2-n+2))+1) \\&+ \frac{2m^3+3m^2+m}{6}\cdot (5n^2-3n) + m\cdot ((5n^2 + 2n - 1)+\\&\lceil\log (w+1)\rceil\cdot(5n^2-n+2)) + m\cdot \frac{3n^2-n-2}{2} +m \\&+ m\cdot \frac{5n^2-5n+4}{2} + \frac{m^2+3m}{2}\cdot (5n^2-3n) + \frac{m^2-m}{2}\\&\cdot\frac{3n^2-3n}{2} + \frac{2m^3+3m^2+m}{6}\cdot \frac{7n^2-3n}{2}\,,\\
    % =&\,,
\end{align*}
\end{minipage}
}

\resizebox{.9\linewidth}{!}{
  \begin{minipage}{\linewidth}
\begin{align*}
    R_{\mathtt{SecRowEch}}(n,m,w) =& \frac{m^2-m}{2}\cdot(R_{\mathtt{SecNonzero}}(n,w)+R_{\mathtt{SecNOT}}(n,w)) \\&+ \frac{2m^3+3m^2+m}{6}\cdot R_{\mathtt{SecCondAdd}}(n,1,w) + m\cdot \\& R_{\mathtt{SecNonzero}}(n,w) + m\cdot R_{\mathtt{FullAdd}}(n,w) + m\cdot R_{\mathtt{B2Minv}}(n,w) \\& + \frac{m^2+3m}{2}\cdot R_{\mathtt{SecScalarMult}}(n,1,w) + \frac{m^2-m}{2}\cdot \\& R_{\mathtt{StrongRefresh}}(n,w) + \frac{2m^3+3m^2+m}{6}\cdot R_{\mathtt{SecMultAdd}}(n,1,w)\\
    =& \frac{m^2-m}{2}\cdot \frac{\lceil\log (w+1)\rceil^2-\lceil\log (w+1)\rceil}{2}\cdot(n^2-n) \\&+ \frac{2m^3+3m^2+m}{6}\cdot (n^2-n)w + m\cdot \frac{\lceil\log (w+1)\rceil^2 -\lceil\log (w+1)\rceil}{2}\cdot \\&(n^2-n) + m\cdot \frac{(n^2-n)w}{2} + m\cdot \frac{n^2-n}{2}\cdot w + \frac{m^2+3m}{2}\cdot \\& (n^2-n)w + \frac{m^2-m}{2}\cdot (\frac{n^2-n}{2}\cdot w) + \frac{2m^3+3m^2+m}{6}\cdot \\& \frac{n^2-n}{2}\cdot w\,.\\
    % =&(1/12)(6m^3n^2w-6m^3nw+3m^2n^2w^2+15m^2n^2w-3m^2nw^2\\
    % &-15m^2nw+3mn^2w^2+27mn^2w-3mnw^2-27mnw)\\
    % =&\frac{6}{12}m^3n^2w-\frac{6}{12}m^3nw+\frac{3}{12}m^2n^2w^2+\frac{15}{12}m^2n^2w-\frac{3}{12}m^2nw^2\\
    % &-\frac{15}{12}m^2nw+\frac{3}{12}mn^2w^2+\frac{27}{12}mn^2w-\frac{3}{12}mnw^2-\frac{27}{12}mnw\\
    % =& \,.
\end{align*}
\end{minipage}
}

\FloatBarrier
\section{Security Proofs of Step 1-4 of Algorithm \ref{alg:rowech}}\label{sec:appen4}
\begin{figure}
    \centering
    % \begin{frame}{}
        % \centering
        \scalebox{0.7}{\setlength{\nodedistance}{4mm}
\begin{tikzpicture}[ % <--- used style are moved here 
node distance = \nodedistance,
data/.style = {},
NIop/.style = {rounded corners, draw=black},
SNIop/.style = {double, double distance=0.5mm, rounded corners, draw=black}]

% INPUTS
\node (I_T) [data]  {$(\text{\pmb{T}}_i)$};

\coordinate [right=\nodedistance of I_T] (tmp1);
\coordinate [below=3\nodedistance of tmp1] (tmp2);

% GADGETS
\node (G1_jselect) [NIop, right=\nodedistance of tmp1] {\begin{tabular}{c} $G_1$ \\ \texttt{[j,:]}\end{tabular}};

\node (Tj) [data, right=\nodedistance of G1_jselect]  {$(\text{\pmb{T}}{\scriptstyle [j,:]}_i)$};
\coordinate [right=\nodedistance of Tj] (tmp3);
\coordinate [above=2\nodedistance of tmp3] (tmp4);

\node (G2_kselect) [NIop, right=\nodedistance of tmp2] {\begin{tabular}{c} $G_2$ \\ \texttt{[k,:]}\end{tabular}};

\node (Tk) [data, right=\nodedistance of G2_kselect]  {$(\text{\pmb{T}}{\scriptstyle [k,:]}_i)$};

\node (G3_jjselect) [NIop, right=\nodedistance of tmp3] {\begin{tabular}{c} $G_3$ \\ \texttt{[:,j]}\end{tabular}};

\node (Tjj) [data, right=\nodedistance of G3_jjselect]  {$(\text{\pmb{T}}{\scriptstyle [j,j]}_i)$};

\node (G4_nonzero) [SNIop, right=\nodedistance of Tjj] {\begin{tabular}{c} $G_4$ \\ \texttt{SecNonzero}\end{tabular}};

\node (G5_secnot) [NIop, right=\nodedistance of G4_nonzero] {\begin{tabular}{c} $G_5$ \\ \texttt{SecNOT}\end{tabular}};

\coordinate [right=\nodedistance of G5_secnot] (tmp5);
\coordinate [above=2\nodedistance of tmp5] (tmp6);
\coordinate [below=3\nodedistance of tmp5] (tmp7);

\node (G6_condadd) [SNIop, right=2\nodedistance of G5_secnot] {\begin{tabular}{c} $G_6$ \\ \texttt{SecCondAdd}\end{tabular}};

% OUTPUTS
\node (O_Tj) [data, right=\nodedistance of G6_condadd]  {$(\text{\pmb{T}}{\scriptstyle [j,:]}_i)$};

% CONNECT
\draw [->] (I_T) -- (tmp1) -- (G1_jselect);
\draw [->] (tmp1) -- (tmp2) -- (G2_kselect);
\draw [->] (G1_jselect) -- (Tj);
\draw [->] (G2_kselect) -- (Tk);
\draw [->] (Tj) -- (tmp3) -- (G3_jjselect);
\draw [->] (tmp3) -- (tmp4) -- (tmp6) -- (tmp6 |- G6_condadd.170) -- (G6_condadd.170);
\draw [->] (Tk) -- (tmp7) -- (tmp7 |- G6_condadd.190) -- (G6_condadd.190);
\draw [->] (G3_jjselect) -- (Tjj);
\draw [->] (Tjj) -- (G4_nonzero);
\draw [->] (G4_nonzero) -- (G5_secnot);
\draw [->] (G5_secnot) -- (tmp5) -- (G6_condadd);
\draw [->] (G6_condadd) -- (O_Tj);

\end{tikzpicture}}
    % \end{frame}
    \caption{An abstract diagram of a single iteration $k$ of Step 1 in \texttt{SecRowEch} (Algorithm \ref{alg:rowech}). 
    The $t$-NI gadgets are depicted with a single border and the $t$-SNI gadgets with a double border.
    Probes are defined at the row/column and element level (and not matrix-level) to ensure sound simulation.}
    \label{fig:step1}
\end{figure}
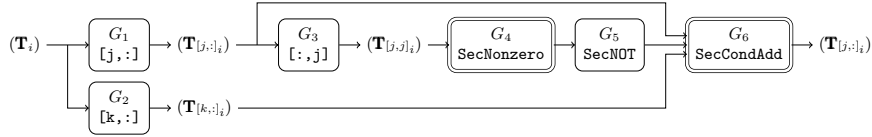

%1: single iteration
%2: total loop -> iterations are sequential
\textbf{\textit{Step 1 ($G_1$ - $G_6$, $t$-SNI, Fig. \ref{fig:step1})}}: we first show that a single iteration of the loop (Line 4-7) is $t$-SNI secure.
We model the extraction of rows $j$ and $k$ (and elements within that row) from matrix $\text{\pmb{T}}$ as $t$-NI gadgets $G_1$ - $G_3$.
This is trivial to show as selected rows from the matrix pass through and the rest is discarded.
The \texttt{SecNonzero} and \texttt{SecNOT} operations are modeled as $t$-SNI and $t$-NI gadgets $G_4$ and $G_5$, respectively.
The \texttt{SecCondAdd} operation is modeled as $t$-SNI gadget $G_6$, operating on rows $j$ and $k$.
An adversary can probe each gadget $G_i$ internally $t_{G_i}$ and at the output $o_{G_i}$. 
The total number of probes in Step 1 is defined as $t_{S_1}$ and the output shares as $|O|$, with
\begin{equation*}
    t_{S_1} = \sum\limits_{i=1}^{6}t_{G_i} + \sum\limits_{i=1}^{5}o_{G_i}, \quad |O| = o_{G_6}
\end{equation*}
We now show that the internal and output probes of each gadget in Step 1 can be perfectly simulated with $\leq t_{S_1}$ input shares.
To simulate the internal and output probes of $G_6$, $t_{G_6}$ shares of each of its inputs are required.
Gadget $G_4$ is $t$-SNI and stops the propagation of probes to the input: only $t_{G_4}$ shares of the output of $G_3$ are required. 
The internal probes and output shares of gadgets $G_1$ and $G_2$ can be simulated with a corresponding number of shares of the input $(\text{\pmb{T}}_i)$, which is a problem as the required input shares is ($t_{G_1} + o_{G_1} + t_{G_2} + o_{G_2} + t_{G_3} + o_{G_3} + t_{G_4} + 2\cdot t_{G_6}$).
It is clear that on a variable-level the probes cannot be perfectly simulated, which is why all gadgets work on a matrix row (or element) level.
As such, each of these gadgets only requires the shares of a specific row of the input. 
And because the rows are independent, $t_{G_6}$ shares of row $k$ and $t_{G_6}$ shares of row $j$ are required, and thus the simulation succeeds.
For the entire composition, the required set of shares of the input is $|I|= t_{G_1} + o_{G_1} + t_{G_2} + o_{G_2} + t_{G_3} + o_{G_3} + t_{G_4} + t_{G_6}$, which is independent from $|O|$.
As a result, each iteration is $t$-SNI secure, and as a result, the whole loop (Step 1) is too. \qed

\begin{figure}
    \centering
    % \begin{frame}{}
        % \centering
        \scalebox{0.7}{\setlength{\nodedistance}{4mm}
\begin{tikzpicture}[ % <--- used style are moved here 
node distance = \nodedistance,
data/.style = {},
NIop/.style = {rounded corners, draw=black},
SNIop/.style = {double, double distance=0.5mm, rounded corners, draw=black}]

% INPUTS
\node (I_Tj) [data]  {$(\text{\pmb{T}}{\scriptstyle [j,:]}_i)$};

% GADGETS
\node (G1_jjselect) [NIop, right=\nodedistance of I_Tj] {\begin{tabular}{c} $G_7$ \\ \texttt{[:,j]}\end{tabular}};

\node (Tjj) [data, right=\nodedistance of G1_jjselect]  {$(\text{\pmb{T}}{\scriptstyle [j,:]}_i)$};

\node (G2_nonzero) [SNIop, right=\nodedistance of Tjj] {\begin{tabular}{c} $G_8$ \\ \texttt{SecNonzero}\end{tabular}};

\node (G3_fulladd) [NIop, right=\nodedistance of G2_nonzero] {\begin{tabular}{c} $G_9$ \\ \texttt{FullAdd}\end{tabular}};

% OUTPUTS
\node (O_cj) [data, right=\nodedistance of G3_fulladd]  {$c{\scriptstyle [j]}$};

% % CONNECT
\draw [->] (I_Tj) -- (G1_jjselect);
\draw [->] (G1_jjselect) -- (Tjj);
\draw [->] (Tjj) -- (G2_nonzero);
\draw [->] (G2_nonzero) -- (G3_fulladd);
\draw [->] (G3_fulladd) -- (O_cj);

\end{tikzpicture}}
    % \end{frame}
    \caption{An abstract diagram of Step 2 in \texttt{SecRowEch} (Algorithm \ref{alg:rowech}). 
    The $t$-NI gadgets are depicted with a single border and the $t$-SNI gadgets with a double border.}
    \label{fig:step2}
\end{figure}
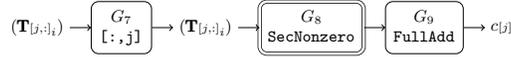

\textbf{\textit{Step 2 ($G_7$ - $G_9$, $t$-NIo, Fig. \ref{fig:step2})}}: we model the extraction of element $j$ from a matrix-row as $t$-NI gadget $G_7$. 
As can be seen in the abstract diagram, \texttt{SecNonzero} is the $t$-SNI gadget $G_8$ and \texttt{FullAdd} is $t$-NI secure $G_9$.
It is clear from its chained structure that the full Step 2 is $t$-NI secure if $\mathbf{c}{\scriptstyle [j]}$ is given to the simulator. 
\qed 
\vspace{-15pt}

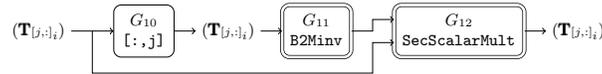
\begin{figure}
    \centering
    % \begin{frame}{}
        % \centering
        \scalebox{0.7}{\setlength{\nodedistance}{4mm}
\begin{tikzpicture}[ % <--- used style are moved here 
node distance = \nodedistance,
data/.style = {},
NIop/.style = {rounded corners, draw=black},
SNIop/.style = {double, double distance=0.5mm, rounded corners, draw=black}]

% INPUTS
\node (I_Tj) [data]  {$(\text{\pmb{T}}{\scriptstyle [j,:]}_i)$};
\coordinate [right=\nodedistance of I_Tj] (tmp1);
\coordinate [below=2\nodedistance of tmp1] (tmp3);

% GADGETS
\node (G1_jjselect) [NIop, right=\nodedistance of tmp1] {\begin{tabular}{c} $G_{10}$ \\ \texttt{[:,j]}\end{tabular}};

\node (Tjj) [data, right=\nodedistance of G1_jjselect]  {$(\text{\pmb{T}}{\scriptstyle [j,:]}_i)$};

\node (G2_b2m) [SNIop, right=\nodedistance of Tjj] {\begin{tabular}{c} $G_{11}$ \\ \texttt{B2Minv}\end{tabular}};

\node (G3_SecScalarMult) [SNIop, right=2\nodedistance of G2_b2m] {\begin{tabular}{c} $G_{12}$ \\ \texttt{SecScalarMult}\end{tabular}};

\coordinate [right=\nodedistance of G2_b2m] (tmp2);
\coordinate [below=2\nodedistance of tmp2] (tmp4);

% OUTPUTS
\node (O_cj) [data, right=\nodedistance of G3_SecScalarMult]  {$(\text{\pmb{T}}{\scriptstyle [j,:]}_i)$};

% % CONNECT
\draw [->] (I_Tj) -- (tmp1) -- (G1_jjselect);
\draw [->] (G1_jjselect) -- (Tjj);
\draw [->] (Tjj) -- (G2_b2m);
\draw [->] (G2_b2m) -- (tmp2) -- (tmp2 |- G3_SecScalarMult.170) -- (G3_SecScalarMult.170);
\draw [->] (G3_SecScalarMult) -- (O_cj);
\draw [->] (tmp1) -- (tmp3) -- (tmp4) -- (tmp4 |- G3_SecScalarMult.190) -- (G3_SecScalarMult.190);

\end{tikzpicture}}
    % \end{frame}
    \caption{An abstract diagram of Step 3 in \texttt{SecRowEch} (Algorithm \ref{alg:rowech}). 
    The $t$-NI gadgets are depicted with a single border, the $t$-SNI gadgets with a double border.}
    \label{fig:step3}
\end{figure}

\textbf{\textit{Step 3 ($G_{10}$ - $G_{12}$, $t$-SNI, Fig. \ref{fig:step3})}}: initially, the pivot is extracted from row $j$ of matrix $(\text{\pmb{T}}_i)$, which we model as $t$-NI gadget $G_{10}$.
The $t$-SNI operations \texttt{B2Minv} and \texttt{SecScalarMult} are modeled as gadgets $G_{11}$ and $G_{12}$, respectively.
An adversary can probe the intermediate values $t_{G_i}$ and output shares $o_{G_i}$ of each gadget $G_i$.
The total number of possible probes in Step 3 is defined as $t_{S_3}$ and its output shares as $|O|$, with
\begin{equation*}
    t_{S_3} = \sum\limits_{i=10}^{12}t_{G_i} + \sum\limits_{i=10}^{11}o_{G_i}, \quad |O| = o_{G_{12}}
\end{equation*}
We now show that the internal and output probes of each gadget in Step 3 can be perfectly simulated with $\leq t_{S_3}$ input shares.
To simulate the $t_{G_{12}}$ intermediate values and $o_{G_{12}}$ output shares of $G_{12}$, only $t_{G_{12}}$ shares of the input $(\text{\pmb{T}}{\scriptstyle [j,:]}_i)$ and the output of $G_{11}$ are required.
For $G_{11}$ too, the $t$-SNI property allows to simulate all probes with only $t_{G_{11}}$ of its input. 
The simulation of $t_{G_{10}} + o_{G_{10}}$ probes on $G_{10}$ requires the same amount of shares from its input.
We now sum the required shares of the input $|I|$ that are required to simulate all probes on the gadgets in Step 3.
As $|I| = t_{G_{10}} + o_{G_{10}} + t_{G_{11}} + t_{G_{12}} \leq t_{S_3}$ and independent of $|O|$, Step 3 is $t$-SNI.
\qed

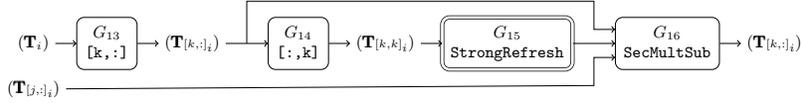
\begin{figure}
    \centering
    % \begin{frame}{}
        % \centering
        \scalebox{0.7}{\setlength{\nodedistance}{4mm}
\begin{tikzpicture}[ % <--- used style are moved here 
node distance = \nodedistance,
data/.style = {},
NIop/.style = {rounded corners, draw=black},
SNIop/.style = {double, double distance=0.5mm, rounded corners, draw=black}]

% INPUTS
\node (I_T) [data]  {$(\text{\pmb{T}}_i)$};
\node (I_Tj) [data, below=0.8\nodedistance of I_T]  {$(\text{\pmb{T}}{\scriptstyle [j,:]}_i)$};

% GADGETS
\node (G1_kselect) [NIop, right=\nodedistance of I_T] {\begin{tabular}{c} $G_{13}$ \\ \texttt{[k,:]}\end{tabular}};

\node (Tk) [data, right=\nodedistance of G1_kselect]  {$(\text{\pmb{T}}{\scriptstyle [k,:]}_i)$};
\coordinate [right=\nodedistance of Tk] (tmp4);
\coordinate [above=2\nodedistance of tmp4] (tmp5);

\node (G2_kkselect) [NIop, right=\nodedistance of tmp4] {\begin{tabular}{c} $G_{14}$ \\ \texttt{[:,k]}\end{tabular}};

\node (Tkk) [data, right=\nodedistance of G2_kkselect]  {$(\text{\pmb{T}}{\scriptstyle [k,k]}_i)$};

\node (G3_refresh) [SNIop, right=\nodedistance of Tkk] {\begin{tabular}{c} $G_{15}$ \\ \texttt{StrongRefresh}\end{tabular}};

\coordinate [right=\nodedistance of G3_refresh] (tmp1);
\coordinate [above=2\nodedistance of tmp1] (tmp2);
\coordinate [below=2\nodedistance of tmp1] (tmp3);

\node (G4_multadd) [NIop, right=\nodedistance of tmp1] {\begin{tabular}{c} $G_{16}$ \\ \texttt{SecMultSub}\end{tabular}};

% OUTPUTS
\node (O_Tk) [data, right=\nodedistance of G4_multadd]  {$(\text{\pmb{T}}{\scriptstyle [k,:]}_i)$};

% CONNECT
\draw [->] (I_T) -- (G1_kselect);
\draw [->] (G1_kselect) -- (Tk);
\draw [->] (Tk) -- (tmp4) -- (G2_kkselect);
\draw [->] (G2_kkselect) -- (Tkk);
\draw [->] (Tkk) -- (G3_refresh);
\draw [->] (G3_refresh) -- (tmp1) -- (G4_multadd);
\draw [->] (I_Tj) -- (tmp3) -- (tmp3 |- G4_multadd.195) -- (G4_multadd.195);
\draw [->] (tmp4) -- (tmp5) -- (tmp2) -- (tmp2 |- G4_multadd.165) -- (G4_multadd.165);
\draw [->] (G4_multadd) -- (O_Tk);

\end{tikzpicture}}
    % \end{frame}
    \caption{An abstract diagram of a single iteration $k$ of Step 4 in \texttt{SecRowEch} (Algorithm \ref{alg:rowech}). 
    The $t$-NI gadgets are depicted with a single border and the $t$-SNI gadgets with a double border.}
    \label{fig:step4}
\end{figure}

\textbf{\textit{Step 4 ($G_{13}$ - $G_{16}$, $t$-NI, Fig. \ref{fig:step4})}}: we first show that a single iteration $k$ of the loop (Line 17-20) is $t$-NI secure.
The extraction of row $k$ and its coefficient $j$ from matrix $(\text{\pmb{T}}_i)$ are modeled as $t$-NI gadgets $G_{13}$ \& $G_{14}$.
We model \texttt{StrongRefresh} as $t$-SNI gadget $G_{15}$ and the $t$-NI secure \texttt{SecMultSub} as $G_{16}$.
An adversary can probe both intermediate values $t_{G_i}$ and output shares $o_{G_i}$ of each gadget $G_i$.
The total number of probes in this step $t_{S_4}$ is defined as:
\begin{equation*}
    t_{S_4} = \sum\limits_{i=13}^{16}t_{G_i} + \sum\limits_{i=13}^{15}o_{G_i}
\end{equation*}
We now show that all probes on intermediate values and output shares of each gadget in Step 4 can be perfectly simulated with $\leq t_{S_4}$ shares of both inputs.
Simulating $t_{G_{16}}$ probes in $G_16$, requires $t_{G_{16}}$ probes of the output of $G_{15}$, input $(\text{\pmb{T}}{\scriptstyle [j,:]}_i)$ and $(\text{\pmb{T}}{\scriptstyle [k,:]}_i)$. 
Gadget $G_{15}$ stops the propagation of probes from the output to the input, as it is $t$-SNI.
Starting at the output of Step 4 and following the flow through all gadgets towards the input, all probes are summed ($|I|$).
As $|I| = t_{G_{13}} + o_{G_{13}} + t_{G_{14}} + o_{G_{14}} + t_{G_{15}} + t_{G_{16}} \leq t_{S_4}$ shares of both inputs are required for simulation of an iteration $k$ in Step 4, it is $t$-NI secure.
As each iteration is independent and computing a single row $k$, they can be assumed to be executed in parallel. 
As a result, we can summarize the gadgets in each iteration as single gadgets across all iterations. 
This means the entire loop is $t$-NI.
\qed

\FloatBarrier

\end{document}